\documentclass[aps,prd,preprint,eqsecnum,tightenlines,nofootinbib,floatfix]{revtex4}
\usepackage[dvips]{graphicx}
\usepackage{epsf}
\usepackage{amsfonts}
\usepackage{amsmath, amsthm, amssymb}
\usepackage{comment}
\usepackage{subfig}
\usepackage{caption}

\usepackage{color}
\usepackage{soul}
\usepackage[normalem]{ulem}

\def\fig#1{Fig.~{\ref{#1}}}
\def\Fig#1{Fig.~{\ref{#1}}}

\def\eqn#1{Eq.~({\ref{#1}})}
\def\eqns#1#2{Eqs.~({\ref{#1}}) and~({\ref{#2}})} 
\def\sect#1{Section~{\ref{#1}}}

\def\NeqFour{\mathcal{N}=4}
\def\NeqFive{\mathcal{N}=5}

\def\LRHH{{\cal L}_{RH\!H}}
\def\LRRR{{\cal L}_{R^3}}

\def\P{{\rm P}}
\def\NP{{\rm NP}}

\def\K{{\cal K}}
\def\tree{{\rm tree}}

\def\twoloop{{2 \mbox{-} \rm loop}}

\def\pol{\varepsilon}
\def\eps{\epsilon}

\newcommand{\code}[1]{\texttt{#1}}
\newcommand*{\caret}{%
  \begingroup
    \fontencoding{T1}%
    \fontfamily{pcr}
    \selectfont
    \string^%
  \endgroup
}

\newif\ifdraft
\drafttrue
\newif\ifpreprint
\preprinttrue
\def\draftnote#1{{\it #1}}
\def\draftnote#1{\ifdraft{\it #1}\fi}

\begin{document}

\ifpreprint
\hbox{UCLA/15/TEP/102 \hskip 2.2 cm  NORDITA-2015-110 \hskip 2.2 cm CERN-PH-TH-2015-218}
\fi

\title{Double-Copy Constructions and Unitarity Cuts}
 
\author{Zvi~Bern${}^{ab}$, Scott~Davies${}^b$, Josh Nohle${}^{c}$}
\affiliation{
${}^a$Department of Physics, CERN Theory Division, CH-1211 Geneva 23, Switzerland\\ $\null$\\
${}^b$Department of Physics and Astronomy,
 University of California at Los Angeles, Los Angeles, CA 90095, USA \\
 $\null$\\
${}^c$Nordita, KTH Royal Institute of Technology and Stockholm
University, Roslagstullsbacken 23, SE-10691 Stockholm, Sweden \\
$\null$ \\
$\null$ \\
}

\begin{abstract}
The duality between color and kinematics enables the construction of
multiloop gravity integrands directly from corresponding gauge-theory
integrands.  This has led to new nontrivial insights into the
structure of gravity theories, including the discovery of enhanced
ultraviolet cancellations. To continue to gain deeper understandings
and probe these new properties, it is crucial to further improve
techniques for constructing multiloop gravity integrands.  In this
paper, we show by example how one can alleviate difficulties
encountered at the multiloop level by relaxing the color-kinematics
duality conditions to hold manifestly only on unitarity cuts instead
of globally on loop integrands.  As an example, we use a minimal
ansatz to construct an integrand for the two-loop four-point
nonsupersymmetric pure Yang-Mills amplitude in $D$ dimensions that is
compatible with these relaxed color-kinematics duality constraints.
We then immediately obtain a corresponding gravity integrand through
the double-copy procedure.  Comments on ultraviolet divergences are
also included.
\end{abstract}


\maketitle

\section{Introduction}

The duality between color and kinematics~\cite{BCJ,BCJLoop} offers a
practical means for obtaining difficult-to-construct higher-loop
scattering amplitudes in gravity theories. This duality, conjectured
by Carrasco, Johansson and one of the authors (BCJ) to hold to all loop orders,
stipulates that there exist forms of amplitude integrands where the
kinematic numerators of all diagrams satisfy the same algebraic
relations as the color factors.  BCJ duality was first formulated
for adjoint-representation states and has recently been generalized to
also include fundamental-representation states~\cite{GravityGhosts,FundRep}.  
Once the duality is manifest in a
gauge-theory amplitude, corresponding gravity amplitude integrands are obtained
simply by replacing gauge-theory color factors with duality-satisfying
kinematic numerators.  This is known as the ``double-copy''
construction of gravity. It effectively reduces the problem of obtaining multiloop
gravity integrands to the much simpler problem of finding color-kinematics 
duality-satisfying gauge-theory
integrands.  The duality also imposes a rigid structure on gauge-theory loop
integrands that can greatly streamline their construction~\cite{ck4l},
including nonplanar contributions.

At loop level, BCJ color-kinematics duality remains a
conjecture. Moreover, as yet there is no constructive means for
finding forms of the integrands where the duality is
manifest. Instead, higher-loop integrands are generally constructed
case by case using ans\"atze~\cite{BCJLoop,ck4l,JJHenrik}, whose
generalized unitarity cuts are then matched to those of the desired
amplitude via the unitarity
method~\cite{UnitarityMethod,UnitarityDDim}.  There has, however, been
important progress in a variety of directions, including understanding
the underlying group-theoretic structure behind the
duality~\cite{OConnellAlgebras}, explicit constructions of loop
integrands~\cite{BCJLoop,ck4l,JJHenrik,
  JoshOneLoop,OneLoopProgress,HigherLoop}, identifying the duality in
classical solutions including those for black
holes~\cite{OConnelBlackHole}, expanding the range of theories where
the duality applies~\cite{MoreTheories,GravityGhosts,FundRep}, as well
as various other studies~\cite{OtherStudies}.  Recent work based on
twistor string theory and scattering equations also offers a new
avenue for constructing gravity loop amplitudes that manifest the
double-copy structure~\cite{Ambitwistors}.

In this paper, we explore a different strategy for improving our ability
to construct multiloop gravity amplitudes. A generic
problem with using ans\"atze to construct kinematic numerators is they
may not be general enough.  This issue is important for
state-of-the-art calculations: For example, it has proven difficult to
construct integrands for the five-loop four-point amplitude of
$\NeqFour$ super-Yang-Mills theory that manifest BCJ duality between
color and kinematics~\cite{FiveLoopN4YM}. Here we will give a two-loop
example in pure Yang-Mills theory that runs into similar difficulties,
where a seemingly reasonable ansatz is not compatible with both global
BCJ duality and unitarity constraints. By ``global'' BCJ constraints,
we mean the full set of BCJ duality constraints on the integrand.
Instead of expanding the ans\"atze---for instance by allowing for
nonlocalities or abandoning relabelling symmetries---the solution we
adopt here is to relax the BCJ duality constraints so that they hold
only on a spanning set of unitarity cuts. A ``spanning set'' of
unitarity cuts refers to a set of unitarity cuts sufficient for
constructing all terms in the amplitude.  We can then employ the
simpler ansatz while maintaining the key double-copy property, thus
allowing us to obtain corresponding gravity loop integrands directly
from gauge-theory ones.

To demonstrate the usefulness of this approach, we utilize it to
construct the $D$-dimensional two-loop four-gluon integrands of pure
Yang-Mills theory in a form compatible with the double-copy
construction of gravity.  The corresponding gravity amplitude is for a
theory of gravity coupled to a dilaton and an antisymmetric tensor. As
a warm up, we first look at the case of four-dimensional
identical-helicity external gluons. A loop integrand satisfying manifest
global BCJ duality was already given in Ref.~\cite{nonSUSYBCJ}.  We
instead choose to work with an earlier form of the
integrand~\cite{AllPlusQCD} that displays exactly the property that
the duality is not manifest on the integrand but is instead manifest
on a spanning set of generalized unitarity cuts.  For general
external-leg polarizations in $D$ dimensions, there is currently no
known representation of the amplitude where global BCJ duality is
manifest.  We first show that a natural ``minimal ansatz'' is not
compatible with both unitarity and manifest global BCJ constraints on
the integrand.  Enlarging an ansatz can quickly become a losing game
because of the rapid proliferation of possible terms. The minimal
ansatz we use has locality, manifest crossing symmetry and
loop-by-loop power-counting no worse than that of ordinary Feynman
diagrams. By loop-by-loop power-counting, we are referring to the
maximum number of powers of each independent loop momentum that can
appear in the numerator of a given diagram.  We show that once we
relax BCJ duality constraints so that they are manifest only on
generalized unitarity cuts instead of the full integrand, the duality
and unitarity constraints are all compatible.  Having the duality
manifest in the cuts ensures that the gravity integrand constructed
through the double-copy procedure will have the correct gravity
cuts. The Yang-Mills integrand we obtain therefore has the double-copy
property and achieves the goal of immediately giving us a
corresponding gravity integrand with a variety of desirable properties
inherited from the Yang-Mills integrand.

Finding improved means for constructing multiloop gravity integrands
is important for studies of ultraviolet properties of gravity
theories. (For a recent update see Ref.~\cite{FiveLoopN4YM}.)
Explicit calculations show that gravity theories have a softer
ultraviolet behavior than known standard-symmetry considerations
predict.  Certain gravity amplitudes possess ``enhanced
cancellations''~\cite{N5FourLoop}, which are defined to be ultraviolet
cancellations that cannot be exhibited diagram by diagram in any
covariant formalism.  (By a covariant formalism, we mean that the only
kinematic denominators in each diagram are those of standard Feynman
propagators.)  This is a new phenomenon not accounted for by
standard-symmetry considerations. In particular, $\NeqFour$
supergravity~\cite{N4Sugra} at three loops~\cite{N4ThreeLoops} and
$\NeqFive$ supergravity at four loops~\cite{N5FourLoop} are
ultraviolet finite.  No satisfactory standard-symmetry explanation has
been found as yet for these cases, despite some
effort~\cite{BHSSuperSpace}.  By four loops, $\NeqFour$ supergravity
does have an ultraviolet divergence~\cite{N4FourLoops}, but it has a
curious structure connected to a duality-symmetry
anomaly~\cite{N4Anomaly} not present in ${\cal N} \ge 5$
supergravities.  In fact, as recently shown, divergences in gravity
are much more subtle than symmetry considerations suggest: They can be
modified at leading order by evanescent effects and change under
duality transformations~\cite{Evanescent}.  (``Evanescent effects''
refer to effects that vanish strictly in $D=4$ but lead to nontrivial
contributions in dimensional regularization.)  These results emphasize
the need for improved, more powerful methods to study the various
surprising phenomena that become visible only at the multiloop level.
The present paper is a modest step in this direction.

This paper is organized as follows.  In \sect{sec:BCJReview}, we first
review the duality between color and kinematics before describing our
approach of applying it to generalized unitarity cuts.  Then in
\sect{sec:AllPlus}, we present the identical-helicity two-loop
four-point amplitude as a warm-up to the case with general
polarizations, which is subsequently discussed in \sect{sec:Formal}.  In
\sect{sec:Ultraviolet}, we give the two-loop ultraviolet divergence
of the gravity theory, extracted from the identical-helicity amplitude. We
present our conclusions in \sect{sec:Conclusions}.

\section{Duality Between Color and Kinematics}
\label{sec:BCJReview}

In this section, we first review BCJ duality between color and
kinematics~\cite{BCJ,BCJLoop} and then explain our procedure for
imposing it on unitarity cuts.  At loop level, the duality remains a
conjecture, but even so it can greatly streamline the construction of
loop amplitudes~\cite{ck4l} for two reasons:
\begin{enumerate}
\item It imposes a structure on gauge-theory amplitudes that can be
  exploited to determine the full loop integrand from a small subset
  of diagrams called ``master diagrams''. 
\item Once a form of gauge-theory loop integrands has been found
  where the duality is manifest, corresponding gravity integrands are
  easily constructed by replacing color factors with gauge-theory
  numerators that satisfy the duality.
\end{enumerate}
The unitarity method~\cite{UnitarityMethod} then offers a convenient
way to confirm that the construction is correct.  This needs to be
done case by case since BCJ duality remains a conjecture at loop
level.

Here we propose applying a less powerful form of the duality that
retains the above second property at the expense of losing the first
one.  We impose that the duality is manifest on
generalized unitarity cuts that decompose loops into trees and relax
the condition that it be manifest in full loop integrands. This
strategy can be helpful for constructing gravity integrands whenever it
is difficult to find expressions that are compatible with both
unitarity and manifest global BCJ duality in the full integrand.

\subsection{Tree Level}

An $m$-point gauge-theory tree amplitude with all particles in the
adjoint representation can always be written as
\begin{align}
\mathcal{A}^{\tree}_{m}= g^{m-2}\sum_{j}
\frac{c_{j} n_{j}}{\prod_{\alpha_{j}}p^{2}_{\alpha_{j}}}\,,
\label{CubicRepresentation}
\end{align}
where the sum over $j$ is over the set of
distinct $m$-point graphs with only cubic vertices.  Inequivalent relabelings
of a given diagram are counted as distinct graphs, and $g$ is the
gauge coupling constant.  Associated with each graph $j$ are the
following:
\begin{itemize}
\item $1/\prod_{\alpha_{j}} p^{2}_{\alpha_{j}}$: The Feynman propagators
  affiliated with the graph. (The associated factors of $i$ are absorbed
  into the numerators.)

\item $c_{j}$: The color factor obtained by dressing every vertex of
  the graph with the group-theory structure constant,
  $\tilde{f}^{abc}=i\sqrt{2}f^{abc}=\mathrm{Tr}([T^{a},T^{b}] T^{c})$,
  where the hermitian generators of the gauge group are normalized via
  $\mathrm{Tr}(T^{a}T^{b})=\delta^{a b}.$

\item $n_{j}$: The numerator that contains the nontrivial kinematic
  information, including momenta, polarizations and possibly spinors.
\end{itemize}
To obtain the form in \eqn{CubicRepresentation}, we convert
contributions from contact-term diagrams---those with higher-than-three-point vertices---to ones with only cubic
vertices by multiplying and dividing by the appropriate propagators,
i.e., inserting factors of $p_{\alpha}^{2}/p_{\alpha}^{2} =1$.  

\begin{figure}
\includegraphics[scale=.5]{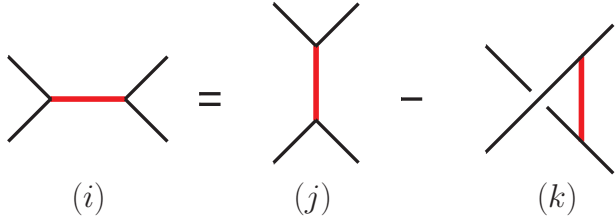}
\caption{The basic Jacobi relation for either color or numerator
  factors.  These three diagrams can be embedded in a larger diagram
  at tree level or loop level. The propagator around which the Jacobi 
  relation is performed is shaded (red).
 }
\label{Fig:BCJ}
\end{figure}

A nontrivial task is to find kinematic numerators that satisfy 
the duality between color and kinematics.
The numerators appearing in \eqn{CubicRepresentation} are by no means
unique due to the freedom to move terms between different
diagrams, also known as generalized gauge
invariance~\cite{BCJ,TyeBCJ,BCJLoop,GravYM2}.  This
freedom can be utilized to find representations of the amplitude where
the kinematic numerators obey the same algebraic relations that the
color factors obey~\cite{BCJ,BCJLoop}.  For adjoint representations in
ordinary gauge theories, this is simply the Jacobi identity,
\begin{equation}
c_{i}=c_{j}-c_{k}\ \Rightarrow\ n_{i}=n_{j}-n_{k} \,,
\label{BCJDuality}
\end{equation}
where $i$, $j$ and $k$ label three diagrams whose color factors obey
the Jacobi identity.  The basic Jacobi relation is displayed in
\fig{Fig:BCJ}.  The generalization of the identity to $m$-point
tree-level amplitudes is seen diagrammatically by embedding
\fig{Fig:BCJ} in larger diagrams, where the other parts of the three
diagrams remain unaltered under the duality.  Furthermore, whenever
the color factor of a diagram is antisymmetric under a swap of legs,
we require that the numerator obey the same antisymmetry,
\begin{equation}
c_{i} \rightarrow - c_{i}\ \Rightarrow\ n_{i} \rightarrow -n_{i} \,.
\label{BCJFlipSymmetry}
\end{equation}
At tree level, numerators that obey the duality for any number of
external legs are known~\cite{TreeAllN}.  The numerator relations are
nontrivial functional relations because they depend on momenta,
polarizations and spinors, as discussed in some detail in
Refs.~\cite{ck4l, JJHenrikReview}.

A central aspect of the duality is the ease with which gravity
amplitude integrands follow from gauge-theory ones once the duality is made
manifest~\cite{BCJ,BCJLoop}: One simply replaces the color factor of a
gauge-theory amplitude with a kinematic numerator, $\tilde{n}$, from a second gauge
theory which has the duality manifest:
\begin{equation}
c_{i}\ \rightarrow\ \tilde{n}_{i}\,.
\label{ColorSubstitution}
\end{equation}
With this replacement in \eqn{CubicRepresentation}, we obtain the
double-copy form of gravity
tree amplitudes,
\begin{align}
\mathcal{M}^{\tree}_{m} =
i \left(\frac{\kappa}{2}\right)^{m-2}
\sum_j \frac{\tilde{n}_{j}n_{j}}{\prod_{\alpha_{j}}p^{2}_{\alpha_{j}}} \,,
\label{DoubleCopy}
\end{align}
where $\tilde{n}_j$ and $n_j$ are gauge-theory numerator factors and
the gravitatonal coupling is given in terms of Newton's constant via
$\kappa^2 = 32 \pi G_N$.  A
tree-level proof for this construction is given in Ref.~\cite{GravYM2}.  We note that in the
double-copy procedure, only one of the two sets of numerators needs to
satisfy the duality of \eqn{BCJDuality} ~\cite{BCJLoop,GravYM2}. The
specific gravity theory one obtains depends on which gauge
theories are used as inputs into the construction.

\subsection{Loop Level}
\label{sec:LoopLevelBCJ}

At loop level, the duality (\ref{BCJDuality}) remains a
conjecture~\cite{BCJLoop}. As at tree level, we express the amplitudes
as a sum over diagrams with only cubic vertices:
\begin{equation}
\mathcal{A}^{L\hbox{-}\mathrm{loop}}_{m}=i^{L} g^{m-2+2 L}
\sum_{\mathcal{S}_{m}}\sum_{j}\int\prod_{l=1}^{L}\frac{d^{D}p_{l}}{(2\pi)^{D}}
\frac{1}{S_{j}}\frac{c_{j} n_{j}}{\prod_{\alpha_{j}}p^{2}_{\alpha_{j}}}\,.
\label{CubicRepresentationLoop}
\end{equation}
The first sum runs over the $m!$ permutations of the external legs,
denoted by $\mathcal{S}_{m}$.  The $S_{j}$ symmetry factor removes any
overcounting from these permutations and also from any internal
automorphism symmetries of graph $j$. Here the $j$-sum runs over the
set of distinct, nonisomorphic, $L$-loop $m$-point graphs with only
cubic or trivalent vertices.  Again, absorbing numerators of contact diagrams
that contain higher-than-three-point vertices into
numerators of diagrams with only trivalent vertices is trivial.
However, it is nontrivial to make a rearrangement into the BCJ-conjectured 
form, where the numerator factors
obey the same Jacobi relations (\ref{BCJDuality}) and symmetry
properties (\ref{BCJFlipSymmetry}) as the color factors.

The generalization of BCJ duality to loop-level amplitudes is to
embed \fig{Fig:BCJ} in larger loop diagrams~\cite{BCJLoop}.  We then
demand that the numerators of all diagrams obey the BCJ relations
(\ref{BCJDuality}) and (\ref{BCJFlipSymmetry}).  We refer to this as
``global BCJ duality''.  Using the kinematic Jacobi relations, one can
solve for the numerators of all diagrams in terms of a relatively
small number of ``master'' numerators~\cite{ck4l}.

Once we have gauge-theory numerator factors that satisfy the duality, the
substitution of color factors by the numerator factors
(\ref{ColorSubstitution}) gives us the double-copy form of gravity
loop integrands,
\begin{align}
\mathcal{M}^{L\hbox{-}\mathrm{loop}}_{m} = 
i^{L+1} \left(\frac{\kappa}{2}\right)^{m-2+2 L}
\sum_{\mathcal{S}_{m}}\sum_{j}\int\prod_{l=1}^{L}\frac{d^{D}p_{l}}{(2\pi)^{D}}
\frac{1}{S_{j}}\frac{\tilde{n}_{j}n_{j}}{\prod_{\alpha_{j}}p^{2}_{\alpha_{j}}} \,,
\label{DoubleCopyLoop}
\end{align}
where $\tilde{n}_j$ and $n_j$ are gauge-theory numerator factors.  
As at tree level, the theories to which the gravity amplitudes belong are
dictated by the input gauge theories. The gravity theory corresponding
to two copies---call them ``left'', $L$, and ``right'', $R$, copies---of 
nonsupersymmetric pure Yang-Mills theory includes
a graviton, a dilaton and an antisymmetric tensor field. We can see this
by decomposing the product of two gluon polarization vectors into irreducible
parts: a symmetric and traceless term, an antisymmetric term and a trace term.
For instance, for a $D$-dimensional external leg with momentum $k$
and polarization vectors $\pol_{L}$, $\pol_{R}$ with reference 
momentum $q$, we can write
\begin{align}
\pol_{L}^{\mu}\pol_{R}^{\nu}
\,\,=\,\,&
\bigg[\frac{1}{2}
\left(
\pol_{L}^{\mu}\pol_{R}^{\nu}+
\pol_{L}^{\nu}\pol_{R}^{\mu}
\right)
- \frac{1}{D-2}
\Bigl({\eta^{\mu\nu}-\frac{k^{\mu}q^{\nu}+k^{\nu}q^{\mu}}{k\cdot q}} \Bigr)
\pol_{L}\cdot\pol_{R} \biggr]
\nonumber \\[.2cm]&
+ \frac{1}{2} (\pol_{L}^{\mu}\pol_{R}^{\nu}-
\pol_{L}^{\nu}\pol_{R}^{\mu} ) +
\frac{1}{D-2}
\Bigl( \eta^{\mu\nu}-\frac{k^{\mu}q^{\nu}+k^{\nu}q^{\mu}}{k\cdot q}\Bigr)
\pol_{L}\cdot\pol_{R} \,.
\end{align}
These three terms correspond to the polarization tensors of a graviton,
an antisymmetric tensor field and a dilaton, respectively. Note that 
$\eta^{\mu\nu}-(k^{\mu}q^{\nu}+k^{\nu}q^{\mu})/(k\cdot q)$
is the usual projector for the  $(D-2)$-dimensional 
space spanned by the polarization vectors orthogonal to both $k$ and $q$.

\begin{figure}[tb]
\includegraphics[scale=.45]{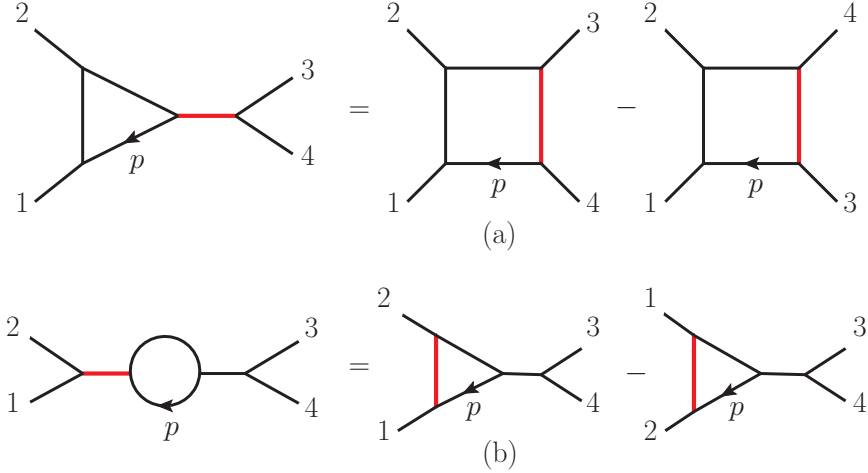}
\caption{The Jacobi relations determining the triangle and bubble 
numerators in terms of box numerators. The shaded (red) propagator
  indicates the line around which the Jacobi identities are
  applied.
}
\label{Fig:TriBub}
\end{figure}

As a simple illustration of color-kinematics duality at loop level,
consider the BCJ numerator identities for one-loop
four-point amplitudes: 
\begin{align}
n^{(1)}_{12(34);p} &= n^{(1)}_{1234;p} - n^{(1)}_{1243;p} \,, \nonumber \\
n^{(1)}_{(12)(34);p} &= n^{(1)}_{12(34);p} - n^{(1)}_{21(34);p} \,, 
\label{eq:snailBCJ}
\end{align}
corresponding to each row of \fig{Fig:TriBub}.  By $n^{(1)}_{1234;p}$ we
mean the box numerator with the external legs following the cyclic
ordering $1234$ and $p$ is the loop momentum.  Similarly,
$n^{(1)}_{(12)34;p}$ and $n^{(1)}_{(12)(34);p}$ denote triangle and
bubble numerators corresponding to the left-most diagrams in
\fig{Fig:TriBub}.
(Here we do not consider the bubble-on-external-leg or tadpole
diagrams that vanish in dimensional regularization after integration.)
Since the numerators of the other diagrams can be
derived from the box diagrams, we call the box diagrams master
diagrams.  If we impose that the numerators obey manifest crossing
symmetry, i.e., that the different box numerators are obtained from
each other simply by appropriate relabelings of the legs, then all
nontrivial information for constructing the amplitude is contained in
a single box numerator.  More generally, at one loop whenever a BCJ
representation of an $m$-point integrand is known, we can construct the
entire amplitude starting from $m$-gon diagrams.

At one loop, there appear to be no difficulties finding numerators
that obey global BCJ identities.  Indeed, there are a variety of known
examples that satisfy global identities in
supersymmetric~\cite{JJHenrik, OneLoopExamples} and
nonsupersymmetric~\cite{nonSUSYBCJ,JoshOneLoop} Yang-Mills theories.

\begin{figure}[tb]
\centering
 \subfloat[]{\includegraphics[scale=.6]{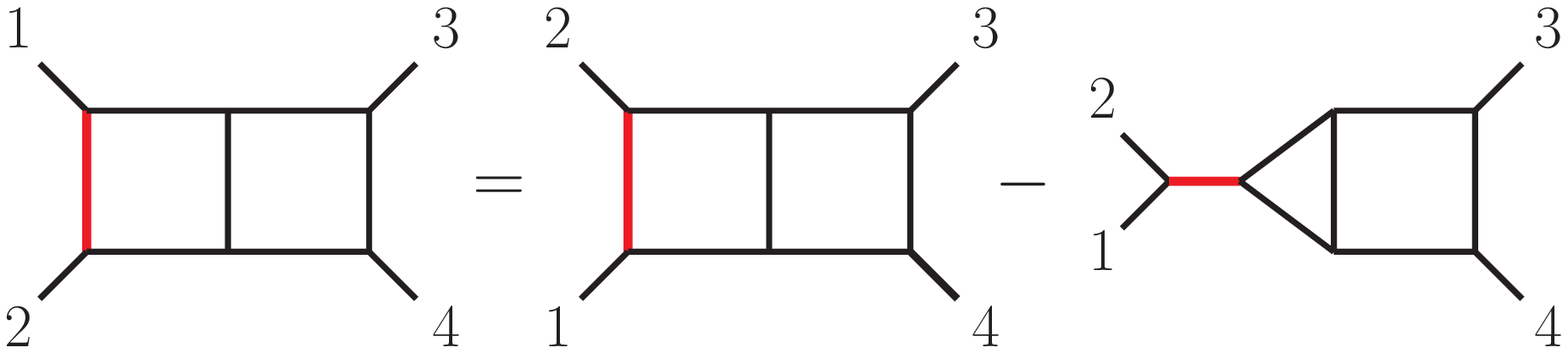}}\\
 \subfloat[]{\includegraphics[scale=.6]{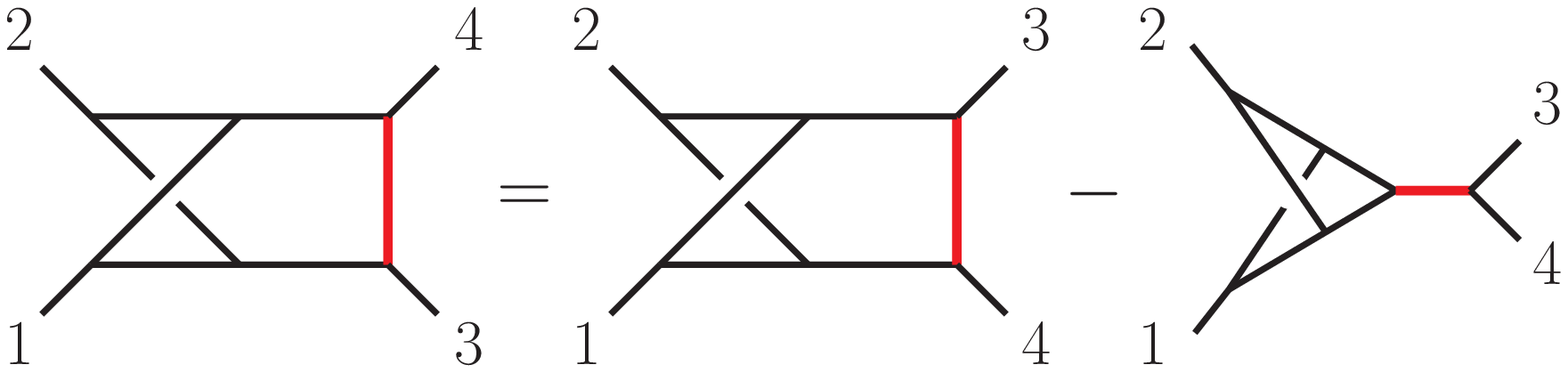}}\\
 \subfloat[]{\includegraphics[scale=.6]{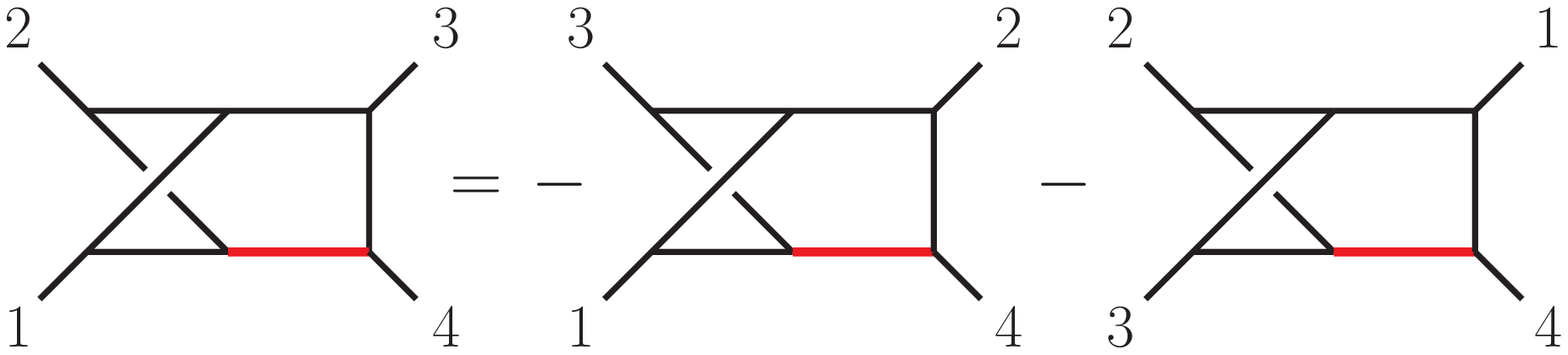}}\\
 \subfloat[]{\includegraphics[scale=.6]{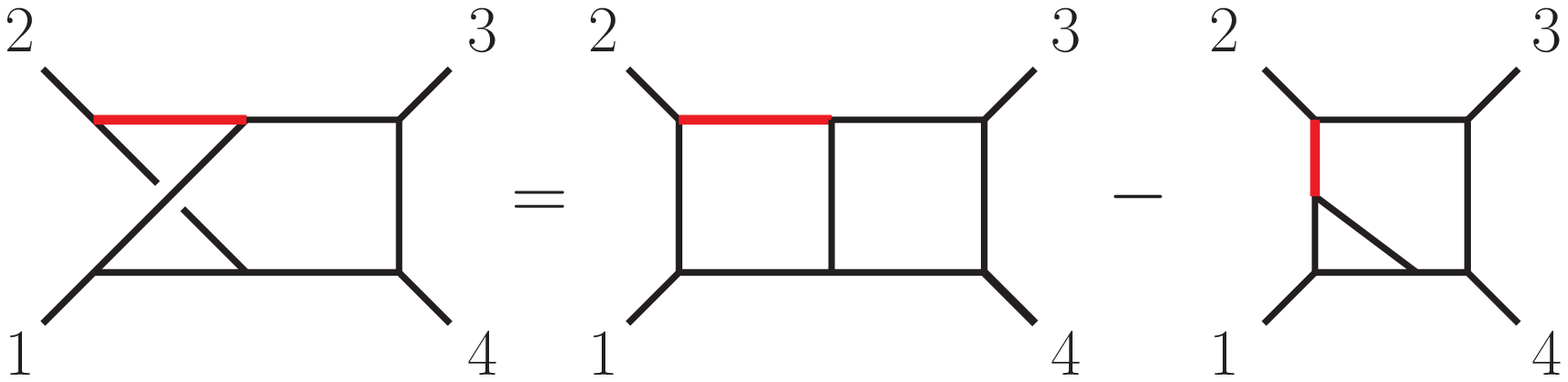}}
\caption[a]{Sample color- or kinematic-numerator Jacobi relations for the
  two-loop four-point amplitudes. The shaded (red) propagator
  indicates the line around which the Jacobi identities are
  applied.
  }
\label{Fig:TwoLoopBCJ}
\end{figure}

At higher loops, the basic ideas are the same although the
constructions rapidly become more difficult.  Nevertheless,
there are a variety of explicit nontrivial examples at two and higher
loops where the global BCJ constraints are manifest~\cite{BCJLoop,ck4l,JJHenrik,MafraSchlottererTwoLoop,nonSUSYBCJ}.
Some examples of duality relations for the two-loop four-point 
amplitude are displayed in \Fig{Fig:TwoLoopBCJ}. 

\subsection{BCJ Duality on Unitarity Cuts}

As the number of loops or legs increases, it becomes more difficult to
find expressions for amplitude integrands where global BCJ
constraints are manifest.  For example, constructing a representation
of the five-loop four-point integrand of $\NeqFour$ super-Yang-Mills
theory with manifest BCJ duality (\ref{BCJDuality})
remains a nontrivial challenge despite its central role for
understanding the ultraviolet properties of gravity theories (see
e.g. Refs.~\cite{FiveLoopN4YM,N5FourLoop}).  Similarly for two-loop
four-point nonsupersymmetric Yang-Mills theory, as we shall see in
\sect{sec:Formal}, a minimal ansatz with local numerators, manifest crossing
symmetry and natural power-counting constraints cannot
simultaneously satisfy global BCJ duality and unitarity.

The obvious strategy is to try to enlarge the ansatz until a solution is
found. However, for nontrivial cases, ans\"atze can grow rapidly,
making them impractical to work with.  Here we take a different
approach: We relax the BCJ duality constraints while keeping the essential
double-copy property, allowing us to obtain
gravity integrands directly from gauge-theory ones.  We find a form of the
amplitude where the BCJ duality relations are manifest in a spanning
set of unitarity cuts rather than on the full uncut integrands. (We note that 
manifest global BCJ duality in the gauge-theory integrand implies that
the BCJ constraints are also valid on the spanning set of unitarity
cuts; however, the converse is not true in general.)  The double-copy
replacement rule of \eqn{ColorSubstitution} then still holds.  In a given
color-dressed Yang-Mills unitarity cut, replacing the color factors by
the corresponding duality-satisfying numerators ensures that the cut
gauge-theory tree integrands composing the cut are properly converted
to cut gravity tree integrands composing the corresponding gravity cut.
If a gauge-theory numerator satisfies the duality in all cuts to which it
contributes, then its double-copy gravity integrand will also satisfy these
cuts and is therefore a valid gravity integrand. Importantly, the requirement that
BCJ duality holds only on the cuts is less restrictive
than having it hold globally in the integrands.  This is true
for two main reasons.  Firstly, the on-shell conditions
can remove contact contributions that violate a given BCJ
numerator Jacobi relation.  Secondly, on a given cut there are fewer
relations because we include only those identities that act separately
on trees composing the cuts.
That is, we do not consider BCJ identities around a cut leg.  It is still
nontrivial, however, because we demand that the duality is manifest
in all cuts of the integrand, which allows use of the double-copy
replacement at the full integrand level.

 \begin{figure}[tb]
     {\includegraphics[scale=.5]{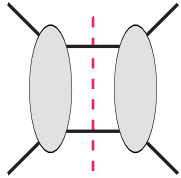}}
\caption{Two-particle cut evaluated in all three channels 
determines one-loop four-point amplitudes.
}
\label{Fig:OneloopCut}
\end{figure}

\begin{figure}[tb]
\includegraphics[scale=.45]{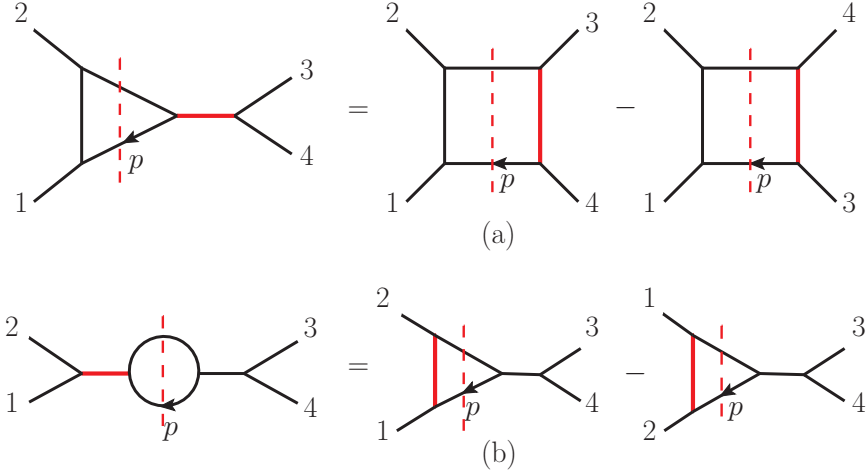}
\caption{The one-loop Jacobi relations with cut conditions
  imposed. The shaded (red) internal lines indicate the leg around which the Jacobi
  identities are applied.  Internal legs intersected by the dashed
  lines are put on shell.  }
\label{Fig:TriBubOnCut}
\end{figure}

 \begin{figure}[tb]
    \subfloat[]{%
     {\includegraphics[scale=.4]{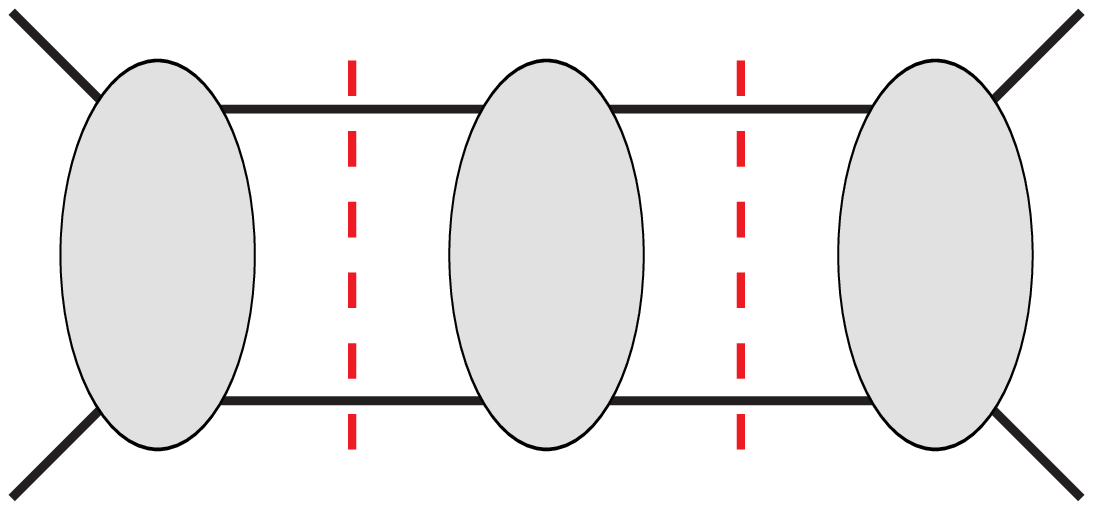}}
    }
    \hspace{1.5cm}
    \subfloat[]{%
       \includegraphics[scale=.4]{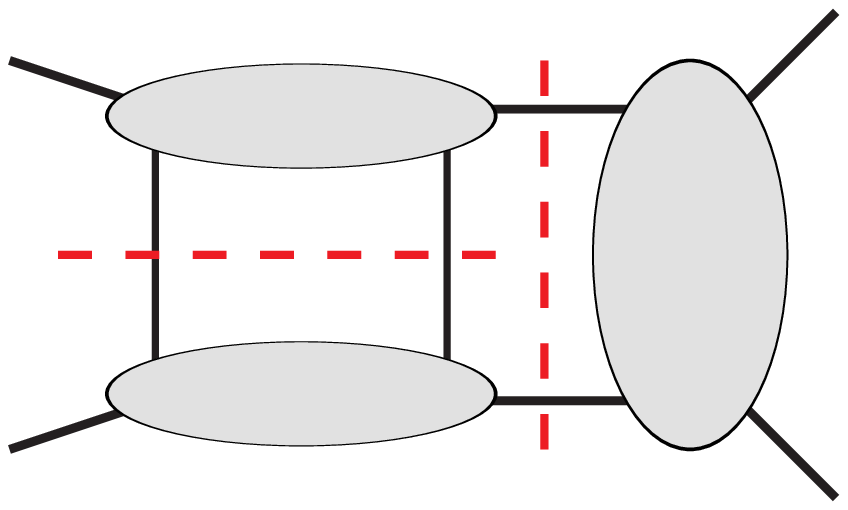}
      }
      \hspace{1.5cm}
    \subfloat[]{%
       \includegraphics[scale=.4]{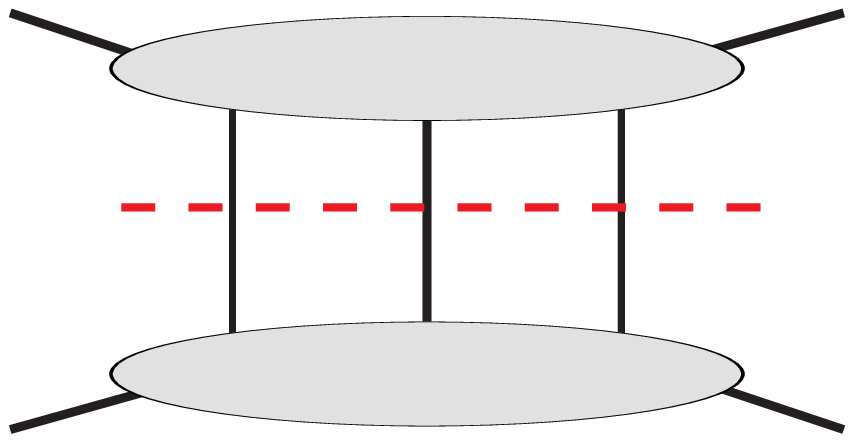}
      }
\caption{A spanning set of generalized unitarity cuts for two-loop four-point color-dressed
gauge-theory or gravity amplitudes.  The exposed internal propagators are put on shell.
}
\label{Fig:UnitarityCuts}
\end{figure}

As a simple first example, consider the one-loop four-point amplitude.
A spanning set of unitarity cuts is the $s$-, $t$- and $u$-channel
versions of the two-particle cut illustrated in \fig{Fig:OneloopCut}.
After enforcing crossing symmetry, we only need to consider the
$s$-channel cut.  Now, instead of imposing the global BCJ conditions
of \eqn{eq:snailBCJ}, as illustrated in \fig{Fig:TriBubOnCut}, we
instead impose
\begin{align}
\label{eq:BCJOnCuts1L}
& \left(n^{(1)}_{12(34);p} - n^{(1)}_{1234;p} + n^{(1)}_{1243;p}\right) 
\Bigr|_{\ell_1^2 = \ell_2^2 = 0}
= 0\,, \nonumber \\
& \left(n^{(1)}_{(12)(34);p} - n^{(1)}_{12(34);p} + n^{(1)}_{21(34);p}\right) 
\Bigr|_{\ell_1^2 = \ell_2^2 = 0}
= 0 \,, 
\end{align}
where $\ell_1 = p$ and $\ell_2 = p-k_1 - k_2$. Whereas
\eqn{eq:snailBCJ} could be thought of as fully defining the
triangle and bubble numerators in terms of the box-numerator master ansatz,
\eqn{eq:BCJOnCuts1L} should be thought of as constraint equations
on separate ans\"atze for the box, triangle and bubble numerators.
Of course, for one-loop four-point amplitudes, there is not
much point in imposing the relaxed BCJ conditions since no difficulties
are encountered when imposing global BCJ constraints directly on the 
integrand. Indeed, there are a number of constructions of such 
integrands including the $D$-dimensional case with formal 
polarizations~\cite{OneLoopExamples, nonSUSYBCJ, JoshOneLoop}.

\begin{figure}[tb]
\centering
\subfloat[]{\hskip -.1 cm \includegraphics[scale=.6]{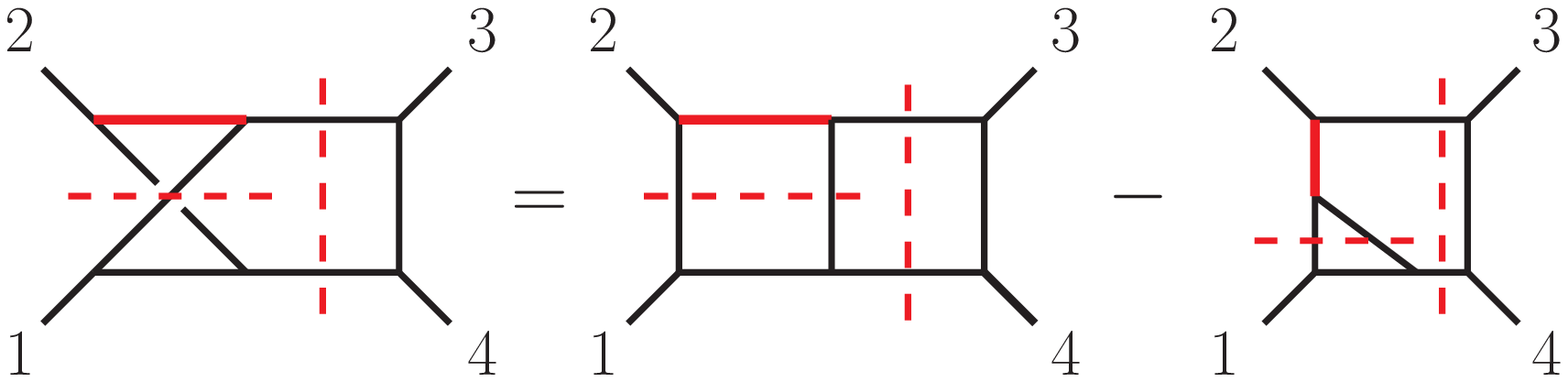}}
\\[.2cm]
\subfloat[]{\hskip -.1 cm \includegraphics[scale=.6]{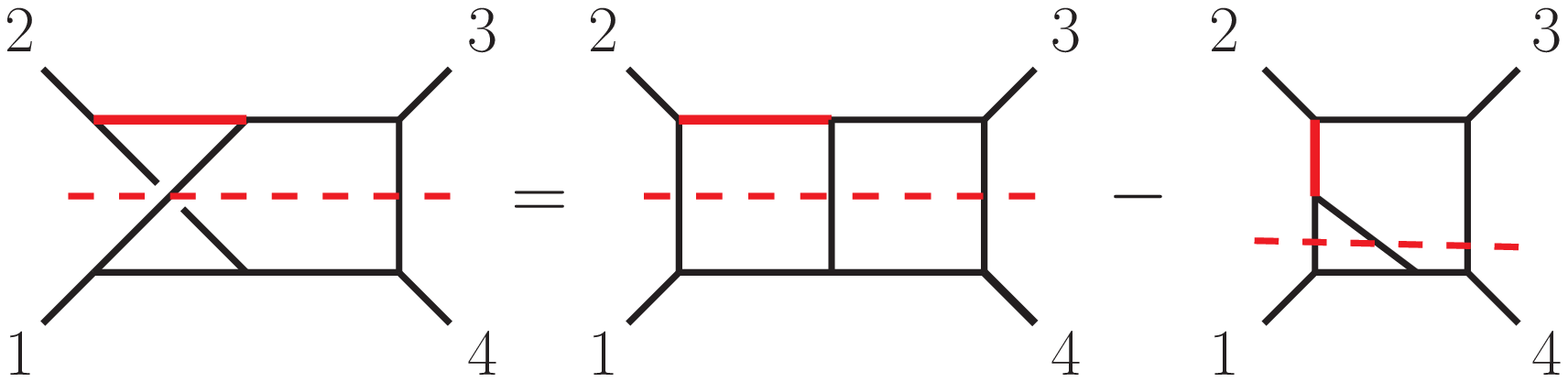}}
\\[.2cm]
\subfloat[]{\hskip .6 cm \includegraphics[scale=.6]{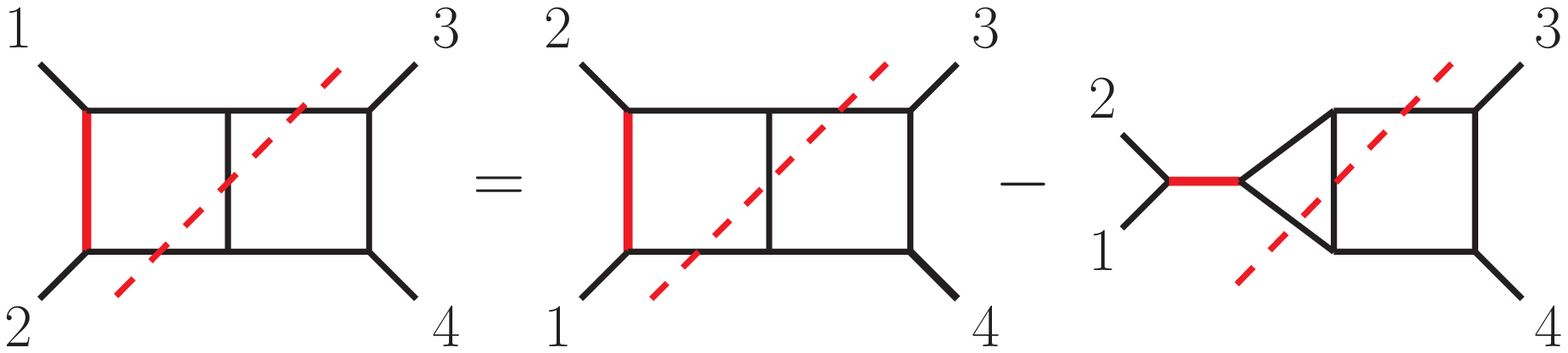}}
\\[.2cm]
\subfloat[]{\hskip .6 cm \includegraphics[scale=.6]{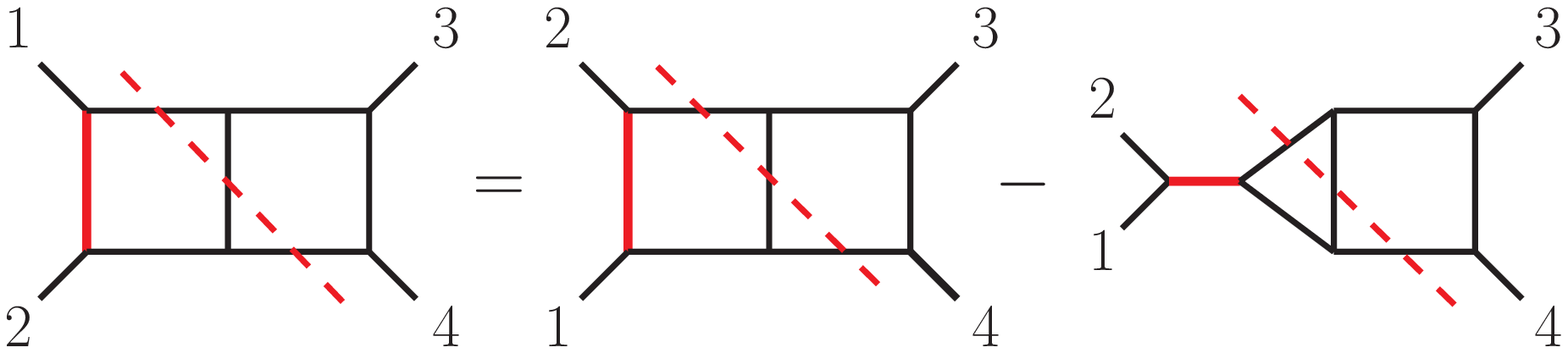}}
\caption{Sample BCJ relations on cut two-loop diagrams.
The diagrams represent kinematic numerators. Internal legs intersected by
the dashed lines are put on shell. }
\label{Fig:TwoLoopCutBCJ}
\end{figure}

At two loops, the basic idea is the same.  \Fig{Fig:UnitarityCuts}
gives a spanning set of unitarity cuts which decomposes the integrand
into sums of products of tree amplitudes.  This set consists of the
iterated two-particle cuts, (a) and (b), and the three-particle cut,
(c).  To impose the BCJ constraints on the cuts, we start with the BCJ
identities on the numerators, as illustrated in \fig{Fig:TwoLoopBCJ}.
We then impose cut conditions as illustrated in
\fig{Fig:TwoLoopCutBCJ}. The figure shows a sample of BCJ relations in an
iterated two-particle cut as well as in three-particle cuts.  Examples
(a) and (b) in \fig{Fig:TwoLoopCutBCJ} are simply the BCJ identity in
\fig{Fig:TwoLoopBCJ}(d) but with on-shell conditions imposed on the
cut legs.  Examples (c) and (d) are slightly trickier because both are
part of the same three-particle cut.  The rule for grouping the cut
diagrams into BCJ triplets is that the three cut diagrams are
identical, including which legs are cut, except for the legs involving
the BCJ duality.   Again, the effect of the cut is to drop 
terms that cancel the cut propagators.  This can help by allowing
the use of a smaller ansatz than would have otherwise been possible.

One practical way to carry out this construction is to start with
kinematic-numerator ans\"atze with free parameters multiplying each
possible term, subject to various desirable properties such as
manifest locality, power-counting constraints and crossing symmetries.
We then solve for some of the free parameters by matching to unitarity
cuts---guaranteeing that the kinematic numerators will produce the
correct Yang-Mills amplitude---and by imposing that BCJ duality holds
on all of the cuts, as illustrated in \fig{Fig:TwoLoopCutBCJ}.
Assuming a consistent solution is found, the so-constructed
gauge-theory integrands give double-copy gravity integrands, as
desired, by replacing the color factors with corresponding numerator
factors.  Because the Yang-Mills integrands obey BCJ duality on each
cut, the double-copy construction leads to gravity integrands that
have the correct unitarity cuts.  We then have correct gravity
integrands, guaranteed by the $D$-dimensional unitarity method.

\section{BCJ Numerator Construction: Two-Loop Four-Point Identical Helicity}
\label{sec:AllPlus}

We consider nonsupersymmetric pure Yang-Mills theory defined by the usual Lagrangian,
\begin{align}
\mathcal{L}_{\text{YM}} = - \frac{1}{4} F^a_{\mu\nu}  F^{a\mu\nu}\,,
\label{eq:LagrangianYM}
\end{align}
where $F^a_{\mu\nu} = \partial_\mu A_\nu^a - \partial_\nu A_\mu^a + g
f^{abc} A^b_\mu A^c_\nu$ is the field strength.  The corresponding
double-copy theory contains a graviton, dilaton and
antisymmetric tensor field, as we discussed
in \sect{sec:LoopLevelBCJ}. The Lagrangian for this theory is
\begin{align}
\mathcal{L}_{\text{DC}}=\sqrt{-g}\left(-\frac{2}{\kappa^2}R
+\frac{1}{2}\partial_{\mu}\phi\partial^{\mu}\phi
+\frac{1}{6} e^{-2\kappa \phi/\sqrt{D-2}} 
H_{\mu\nu\rho}H^{\mu\nu\rho}\right)\,,
\label{eq:LagrangianGDA}
\end{align}
where $H_{\mu\nu\rho}=\partial_{\mu}A_{\nu\rho} +
\partial_{\nu}A_{\rho\mu} + \partial_{\rho}A_{\mu\nu}$, and
$A_{\mu\nu}=-A_{\nu\mu}$ is the rank-two antisymmetric tensor
field. We use the mostly-minus metric convention and the
Ricci-curvature convention, $R_{\mu\nu}\equiv
R^{\alpha}_{\phantom{\alpha}\mu\alpha\nu}$.
Eq.~\eqref{eq:LagrangianGDA} corresponds to the low-energy effective
Lagrangian of the bosonic part of string theory. By using the
double-copy construction, we reduce the problem of constructing
integrands for amplitudes in the theory described by the Lagrangian
(\ref{eq:LagrangianGDA}) to constructing those for Yang-Mills theory
(\ref{eq:LagrangianYM}).  Pure gravity has been considered in
Ref.~\cite{Evanescent}. The construction used there is different due
to the need to introduce explicit physical-state projectors in the
unitarity cuts in order to remove the antisymmetric tensor and dilaton
states from the theory.  A possible route to applying the double-copy
procedure to pure gravity was given in Ref.~\cite{GravityGhosts},
where ghosts are used to cancel unwanted states.

As a warm-up for the two-loop four-point amplitude with arbitrary
$D$-dimensional external states, we first look at the case of
four-dimensional identical-helicity states.  The identical-helicity
pure Yang-Mills amplitude was first constructed in
Ref.~\cite{AllPlusQCD}. The construction relies on a version of the
unitarity method that deals with external four-dimensional helicity
states and internal states in $D$ dimensions, which is appropriate for
dimensional regularization~\cite{UnitarityDDim}.  A full globally
duality-satisfying representation is given in Ref.~\cite{nonSUSYBCJ}.
Here we discuss the original representation from
Ref.~\cite{AllPlusQCD} because it illustrates features we want: It
does not satisfy global BCJ constraints, but it does satisfy the BCJ
identities on the spanning set of cuts in
\fig{Fig:TwoLoopCutBCJ}~\cite{BCJ}.  Therefore, replacing each color
factor with a second copy of the corresponding kinematic numerator
gives the integrand of the double-copy theory.

\begin{figure}[tb]
\centering
 \subfloat[]{\hskip .3 cm \includegraphics[scale=.4]{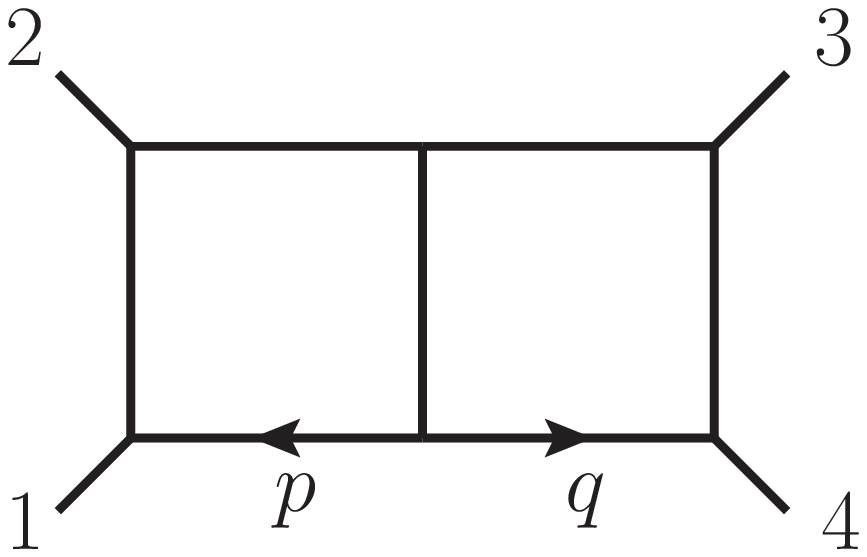}}
\hspace{.5cm}
 \subfloat[]{\hskip .3 cm \includegraphics[scale=.4]{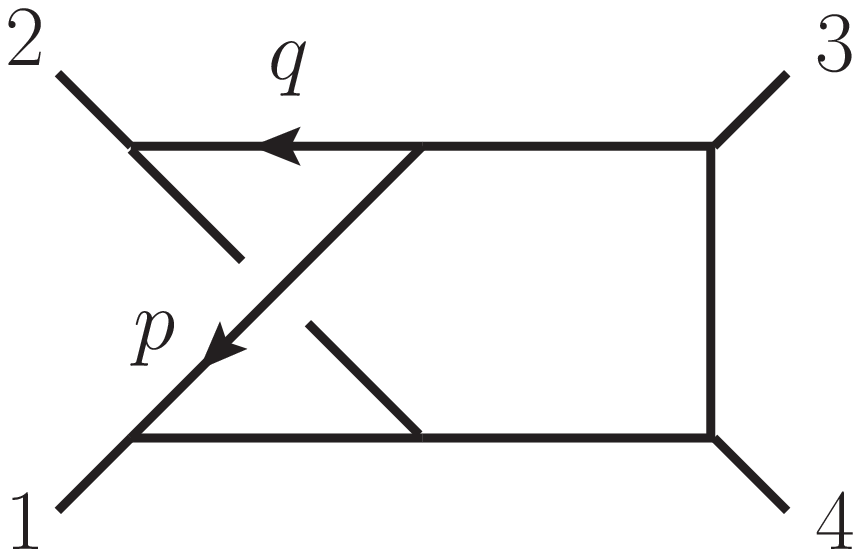}}
\caption[a]{The planar and nonplanar double-box integrals appearing
in the identical-helicity pure Yang-Mills amplitude.
}
\label{Fig:PandNP}
\end{figure}

The two-loop all-plus helicity amplitude for pure Yang-Mills
theory from Ref.~\cite{AllPlusQCD} is
\begin{align}
\mathcal{A}_4^{(2)}(1^+,2^+,3^+,4^+)=
-\frac{g^6}{4}\sum_{\mathcal{S}_4}
\Bigl( c^{\mathrm{P}}_{1234}\, \mathcal{I}^{\P} [n_{1234}^\P]
+ c^{\mathrm{NP}}_{12;34}\,  \mathcal{I}^{\NP}[n_{12;34}^\NP] \Bigr) \,,
\label{eq:twoLoopYM}
\end{align}
where the sum runs over all 24 permutations of the external legs.  The
prefactor $1/4$ accounts for the overcount due to diagram symmetries.
$c^{\mathrm{P}}_{1234}$ and $c^{\mathrm{NP}}_{12;34}$ are the color
factors obtained from the planar and nonplanar double-box diagrams
shown in \fig{Fig:PandNP}(a) and \fig{Fig:PandNP}(b), respectively, 
by dressing each vertex with
an $\tilde{f}^{abc}$ and summing over the contracted color indices.
$\mathcal{I}^{\mathrm{P}}_{1234}$ and
$\mathcal{I}^{\mathrm{NP}}_{12;34}$ are planar and nonplanar
integrals given by
\begin{align}
& \mathcal{I}^{\P}[n_{1234}^\P] = \int\frac{d^Dp}{(2\pi)^D}\frac{d^Dq}{(2\pi)^D} 
\frac{n^\P_{1234}}{p^2q^2(p+q)^2(p-k_1)^2(p-k_1-k_2)^2
(q-k_4)^2(q-k_3-k_4)^2}\,, \nonumber \\
& \mathcal{I}^{\NP}[n_{12;34}^\NP] = \int\frac{d^Dp}{(2\pi)^D} \frac{d^Dq}{(2\pi)^D}
\frac{n^\NP_{12;34}}{p^2q^2(p+q)^2(p-k_1)^2(q-k_2)^2(p+q+k_3)^2
(p+q+k_3+k_4)^2}\,, 
\nonumber \\
\end{align}
with the planar and nonplanar kinematic numerators,
\begin{align}
n_{1234}^\P &=
-i\mathcal{T} \Bigl[
\frac{(D_s-2)^2}{s}(p+q)^2\lambda_p^2\lambda_q^2((p+q)^2+s) 
 + 16s\left((\lambda_p\cdot\lambda_q)^2
 - \lambda_p^2\lambda_q^2\right) \nonumber \\
& \hskip 1 cm \null
+(D_s-2) \Bigl(s\left(\lambda_p^2\lambda_q^2+\lambda_p^2\lambda_{p+q}^2
 + \lambda_q^2\lambda_{p+q}^2\Bigr) 
 + 4 (p+q)^2(\lambda_p^2+\lambda_q^2)(\lambda_p\cdot\lambda_q)\right)\Bigr]\,,
                                         \nonumber \\
n_{12;34}^\NP &=
-i\mathcal{T} s \bigl[(D_s-2)\left(\lambda_p^2\lambda_q^2
  +\lambda_p^2\lambda_{p+q}^2 + \lambda_q^2\lambda_{p+q}^2\right) 
+16\left((\lambda_p\cdot\lambda_q)^2-\lambda_p^2\lambda_q^2\right)\bigr] \,,
\label{eq:PNP}
\end{align}
where $\lambda_p$, $\lambda_q$ and $\lambda_{p+q}$ represent the
$(-2\epsilon)$-dimensional components of loop momenta $p$, $q$ and
$(p+q)$, respectively.  We take $D = 4 -2\eps > 4$ so that $\lambda_p \cdot k_i =
0$, where the $k_i$ are four-dimensional external momenta.  We have
suppressed loop momentum labels for the numerators.  The
state-counting parameter is defined by $D_s = \delta^\mu{}_\mu$ so
that $D_s -2$ corresponds to the number of gluons states for each
color circulating in the loop. The permutation-invariant kinematic
prefactor is given by
\begin{equation}
\mathcal{T}\equiv\frac{[1\,2][3\,4]}
{\langle 1\,2\rangle\langle 3\,4\rangle}\,,
\label{Tdef}
\end{equation}
where $[i\,j]$ and $\langle
i\,j\rangle$ are spinor products, defined in, for example,
Ref.~\cite{ManganoReview}, and $s= (k_1+k_2)^2$ is a
Mandelstam invariant. We have slightly rearranged the form of the
amplitude given in Ref.~\cite{AllPlusQCD} by absorbing the ``bow-tie''
contributions into the planar double box.  With this choice the planar
double-box numerator is nonlocal since it contains a term with a
factor of $1/s$.  In this representation of the amplitude, all
numerators of diagrams with topologies not matching either the planar
or nonplanar double-box diagram are taken to have vanishing
numerators.

Now let us examine the issue of BCJ duality.  As a first example,
consider duality relation (a) in \fig{Fig:TwoLoopBCJ}.  Since all
numerators except the planar and nonplanar double-box ones vanish,
the duality relation reads,
\begin{equation}
n^\P_{2134} = n^\P_{1234} \,.
\end{equation}
In this case, the duality is satisfied trivially, even without any cut
conditions imposed, because the planar numerator in
\eqn{eq:PNP} has a symmetry under the interchange of legs 1 and 2.  The
same holds for the nonplanar diagrams in BCJ relation (b) of 
\fig{Fig:TwoLoopBCJ}.  As a somewhat less trivial example, the third
BCJ relation, (c), is also satisfied. This happens because the nonplanar
numerator, $n^\NP_{12;34}$ of \eqn{eq:PNP}, is independent of external labels
except for the overall factor of $s$. Consequently, the BCJ relation is then
\begin{equation}
n^\NP_{12;34} + n^\NP_{13;24} + n^\NP_{32;14} \propto (s+t+u) = 0 \,,
\end{equation}
where $t = (k_2 + k_3)^2$ and $u = (k_1 + k_3)^2$ are the two
other four-point Mandelstam invariants.

BCJ relation (d) in \fig{Fig:TwoLoopBCJ}, however, does not hold if we do
not impose cut conditions.  Since the ``triangle-in-box'' numerator is zero 
in our representation, the relation reduces to a two-term identity,
\begin{align}
n^\NP_{12;34} = n^\P_{1234}\,.
\label{NPBCJExample}
\end{align}
This identity is obviously not satisfied by the numerators in \eqn{eq:PNP}
since the planar and nonplanar expressions are different.
However, applying either the cut conditions in \fig{Fig:TwoLoopCutBCJ}(a) or
\fig{Fig:TwoLoopCutBCJ}(b) discards all terms proportional to $(p+q)^2$,
thereby removing discrepant terms in \eqn{NPBCJExample}. (Note that for 
this relation we only need to consider the cut conditions of 
\fig{Fig:TwoLoopCutBCJ}(a--b) since the BCJ 
relation of \fig{Fig:TwoLoopBCJ}(d) does not appear in cut (a) of 
\fig{Fig:UnitarityCuts}.)

By systematically proceeding through all of the cuts in
\fig{Fig:UnitarityCuts}, it is straightforward to check that all BCJ
identities hold for all diagrams composing each cut.  This implies
that a double-copy gravity integrand is obtained simply by replacing
the color factors in \eqn{eq:twoLoopYM} with a kinematic numerator and
replacing the gauge-theory coupling with the gravitational one.  Thus
we obtain an expression for the identical-helicity two-loop
four-graviton amplitude in the double-copy theory
(\ref{eq:LagrangianGDA}),
\begin{align}
\mathcal{M}_4^{(2)}(1^+,2^+,3^+,4^+)= -i
\Bigl(\frac{\kappa}{2}\Bigr)^6 \frac{1}{4}\sum_{\mathcal{S}_4}
\Bigl(\mathcal{I}^{\P}[(n^\P_{1234})^2] + \mathcal{I}^{\NP}[(n^\NP_{12;34})^2] \Bigr)\,.
\label{eq:twoLoopGR}
\end{align}
The kinematic numerators are the squares of the gauge-theory ones in
\eqn{eq:PNP}.  We have directly confirmed that the spanning set of
unitarity cuts (\fig{Fig:UnitarityCuts}) of \eqn{eq:twoLoopGR} are
 all correct, where the internal legs are taken to 
be in $D$ dimensions.

 \begin{figure}[p]
\renewcommand*\thesubfigure{\arabic{subfigure}}
     \centering
    \subfloat[]{%
      {\hskip .6 cm \includegraphics[scale=.45]{diag1.eps}}
    }
    \hspace{.5cm}
    \subfloat[]{%
      {\hskip .6 cm \includegraphics[scale=.45]{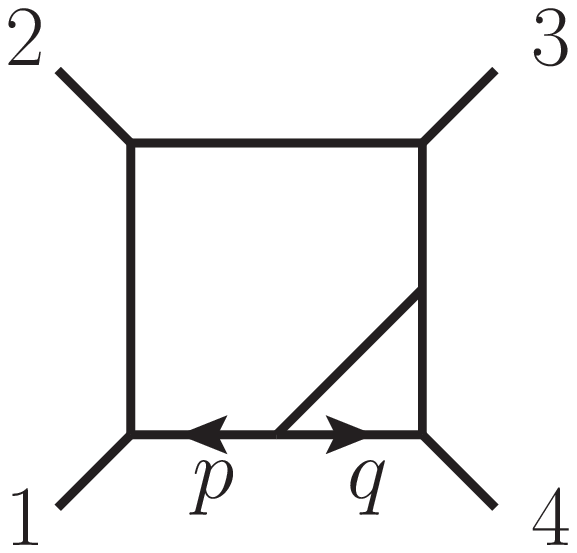}}
    }
    \hspace{.5cm}
    \subfloat[]{%
      {\hskip .4 cm \includegraphics[scale=.45]{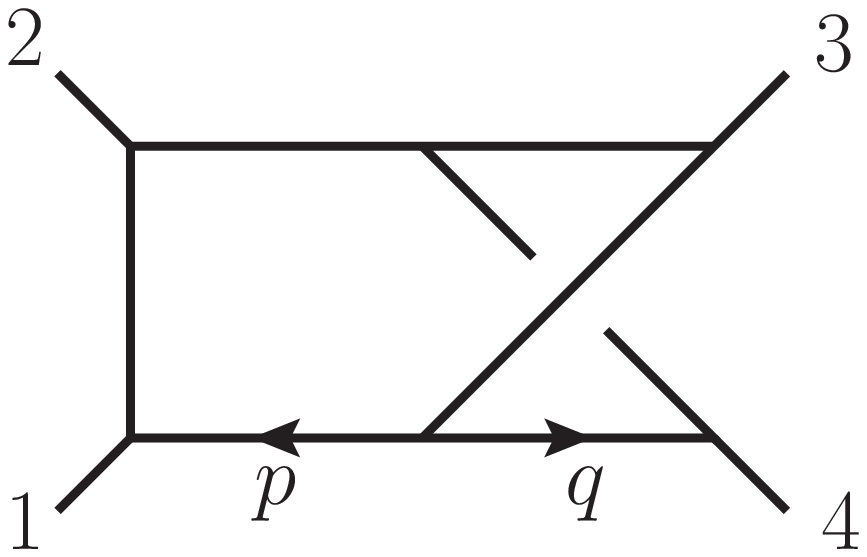}}
    }
    \\[.2cm]
    \hskip -.4 cm 
    \subfloat[]{%
     {\hskip .4 cm  \includegraphics[scale=.45]{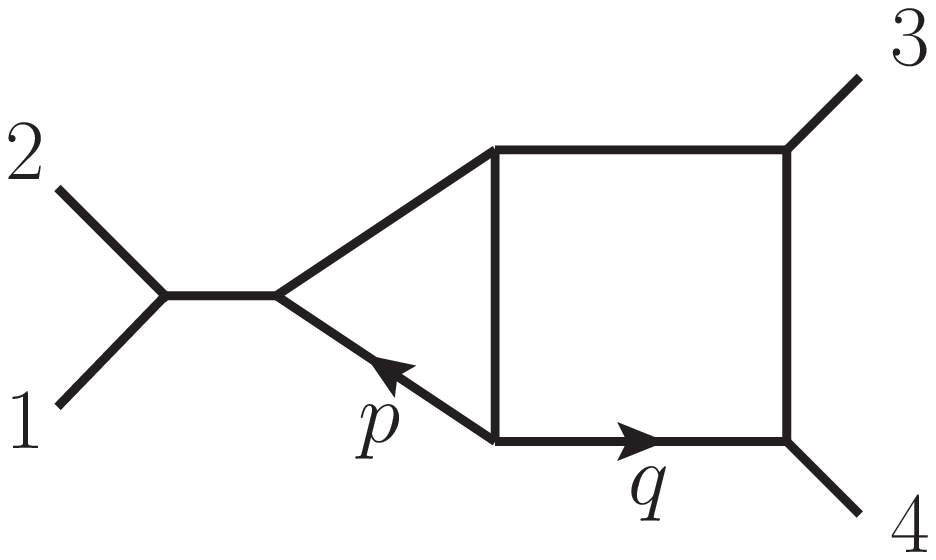}}
    }
    \hspace{.4cm}
    \subfloat[]{%
     {\hskip .4 cm   \includegraphics[scale=.45]{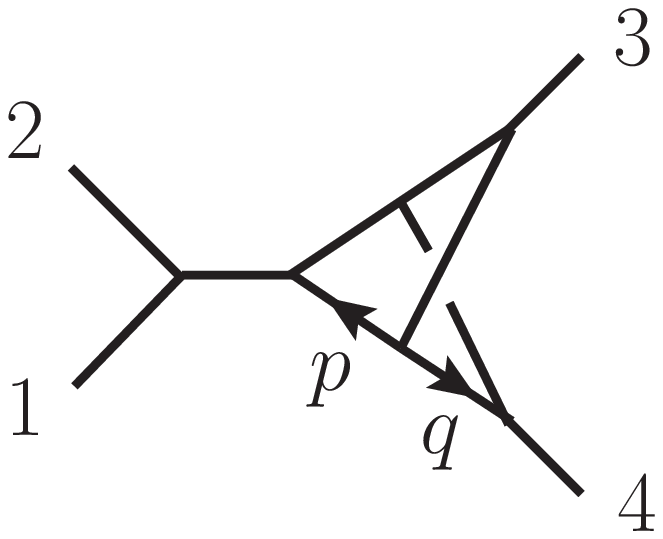}}
    }
    \hspace{.6cm}
    \subfloat[]{%
      {\hskip .4 cm \includegraphics[scale=.45]{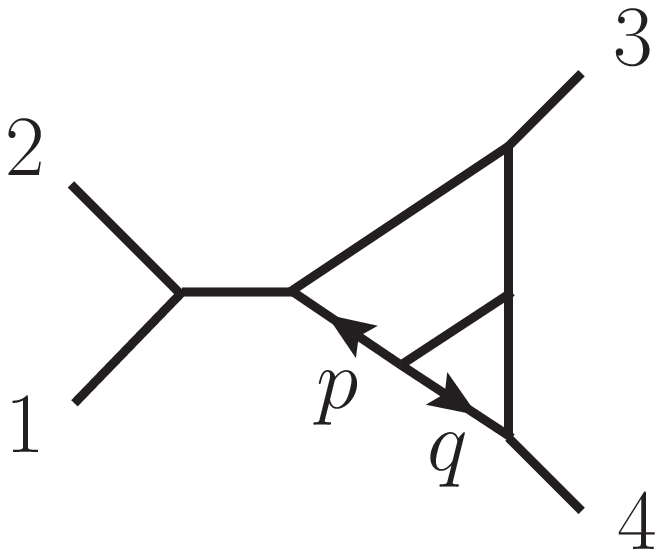}}
    }
    \\[.2cm]
    \subfloat[]{%
     {\hskip .4 cm \includegraphics[scale=.45]{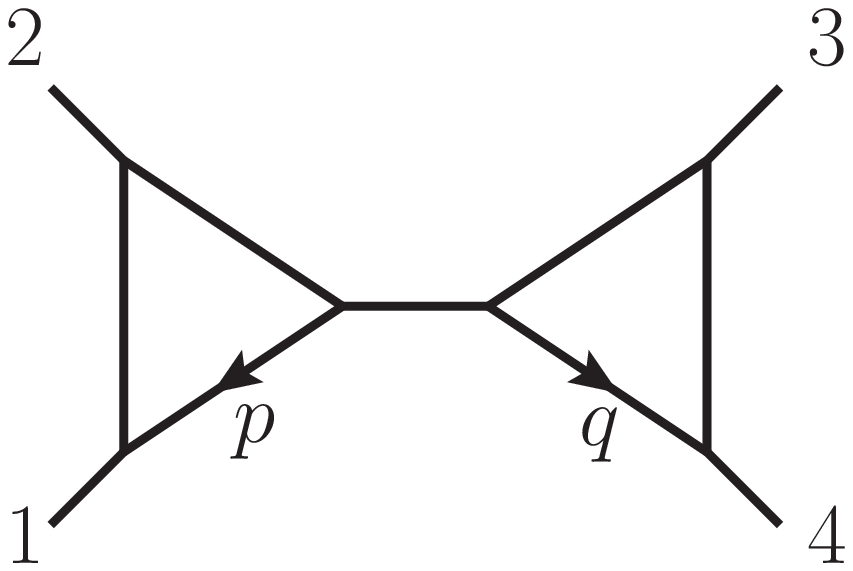}}
    }
    \hspace{.1cm}
    \subfloat[]{%
     {\hskip .6 cm \includegraphics[scale=.45]{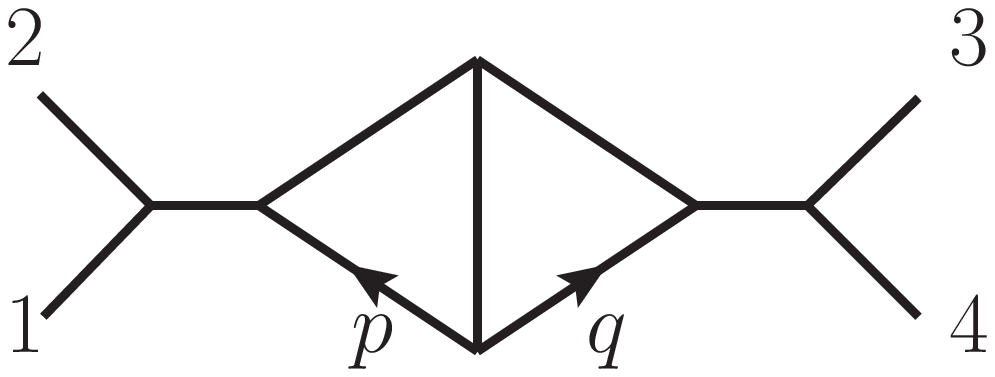}}
    }
    \hspace{-.2cm}
    \subfloat[]{%
    { \hskip .5 cm \includegraphics[scale=.45]{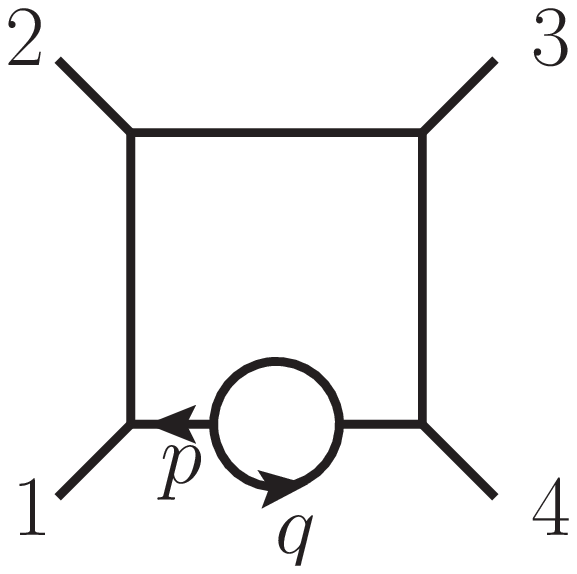}}
    }
    \\[.4cm]
    \subfloat[]{%
     {\hskip .5 cm  \includegraphics[scale=.45]{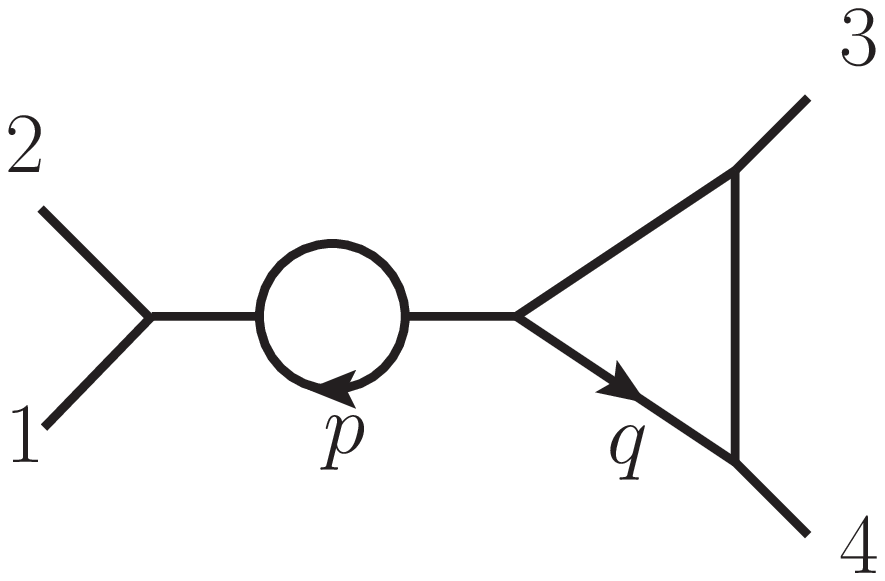}}
    }
    \hspace{.5cm}
    \subfloat[]{%
     {\hskip .5 cm  \includegraphics[scale=.45]{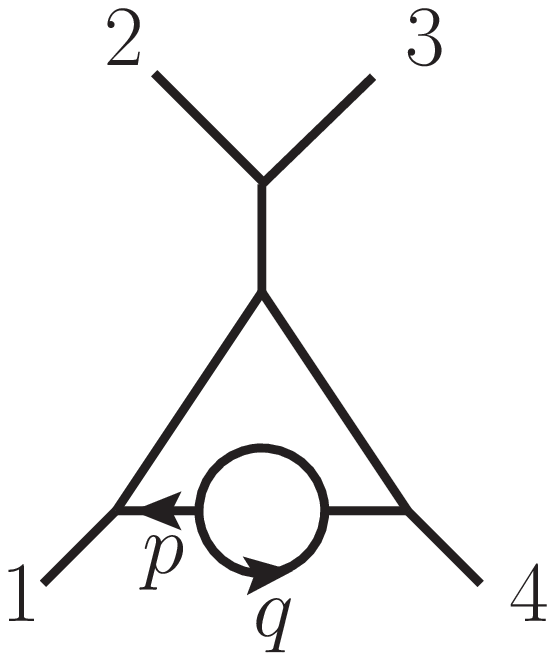}}
    }
    \hspace{.5cm}
    \subfloat[]{%
     {\hskip .4 cm \includegraphics[scale=.45]{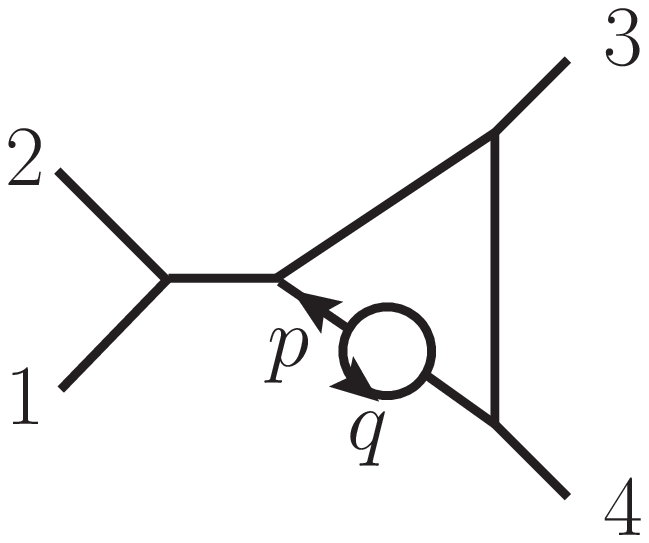}}
    }
    \\[.4cm]
    \subfloat[]{%
     {\hskip .6 cm \includegraphics[scale=.45]{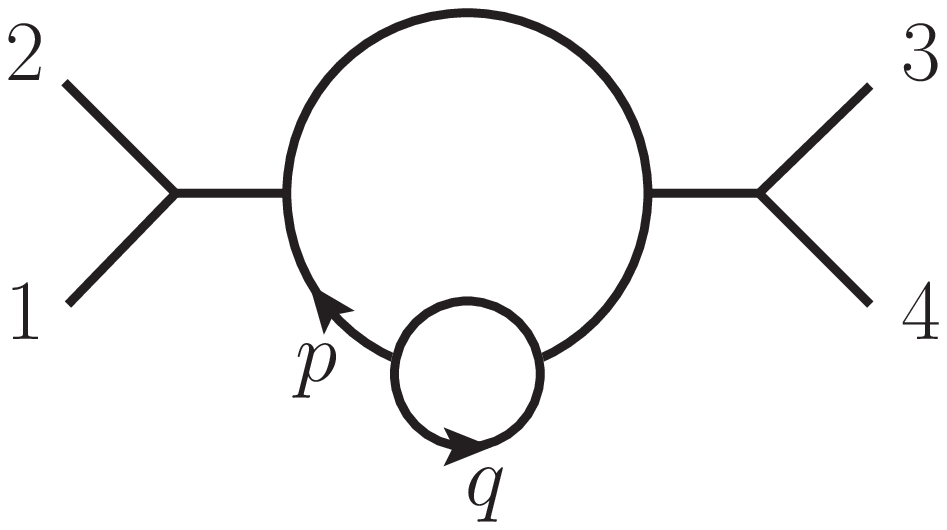}}
    }
    \hspace{.5cm}
    \subfloat[]{%
     {\hskip .5 cm  \includegraphics[scale=.45]{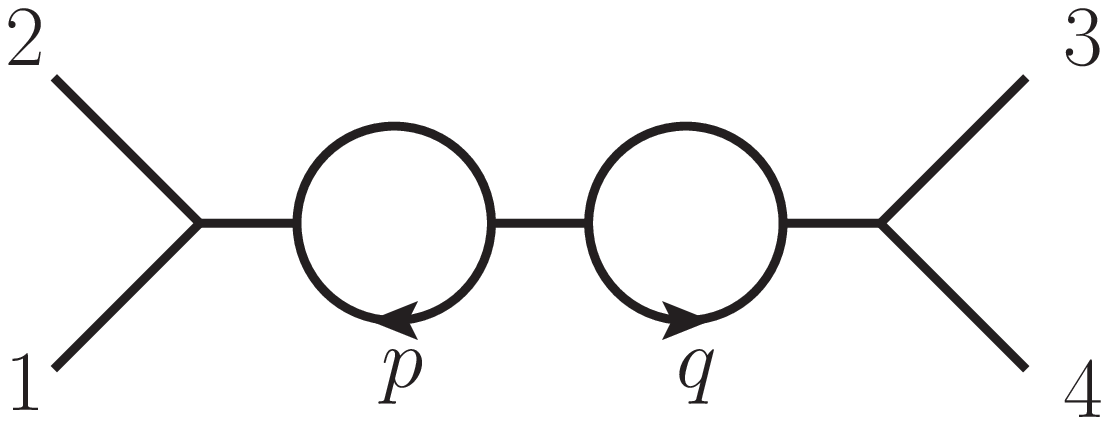}}
    }
    \caption{Two-loop four-point diagrams with only cubic
     vertices. All other diagrams in the amplitude are obtained
     by relabeling external legs.  Tadpole and bubble-on-external-leg
     diagrams, which integrate to zero, are not included.
     }
    \label{Fig:TwoLoopDiags}
  \end{figure}

\section{BCJ Numerator Construction: Two-Loop Four-Point Amplitude 
in $D$ Dimensions}
\label{sec:Formal}

In this section, we describe the construction of the two-loop
four-point double-copy gravity numerators with general formal
polarization tensors in $D$ dimensions.  This example demonstrates how
one can sidestep difficulties when ans\"atze are not sufficiently
general to satisfy global BCJ duality. We do so by loosening the
duality requirements to be manifest only on a spanning set of
generalized unitarity cuts.  Although the use of formal polarization
vectors and tensors leads to much more complicated expressions than
those for helicity amplitudes, it does have the advantage that formal
polarization amplitudes are valid in all dimensions, and the more
straightforward conventional dimensional-regularization
scheme~\cite{CDR} can be used instead of the four-dimensional-helicity
scheme~\cite{FDH}.  They are also useful for
studying evanescent effects in gravity, which can be quite
subtle~\cite{Evanescent}.

\subsection{Constraining an Ansatz}

The construction begins in pure Yang-Mills theory.  Our goal is to
find a representation of the pure Yang-Mills two-loop four-point amplitude of the
form given in \eqn{CubicRepresentationLoop} that has the double-copy
property, allowing us to immediately obtain the
corresponding gravity amplitude in the form of \eqn{DoubleCopyLoop}.

To start, we build an ansatz for the kinematic numerators
with the following properties imposed:
\begin{enumerate}
\item \textit{Locality.} The numerators are local polynomials in
  momenta and polarization vectors, meaning the only allowed kinematic
  denominators in the amplitude are ordinary scalar Feynman
  propagators.
\item \textit{Power-counting.} The same loop-by-loop power-counting
  found from Feynman-gauge Feynman rules is imposed.  For example,
  each term in the numerator of diagram (2) in \fig{Fig:TwoLoopDiags}
  should have a maximum of five powers of $p$, three powers of $q$ and exactly
  six powers total of internal or external momentum. 
\item \textit {Cubic vertices.} The allowed diagrams are ones 
   with only cubic vertices as in \eqn{CubicRepresentationLoop}.
\item \textit{Relabeling.}  The diagrams are functions of
  the external labels.  That is, we can obtain a relabeled diagram's
  numerator simply by relabeling momenta, polarizations and color factors.
\item \textit{Diagram symmetries respected.}  The numerator should
  reflect the symmetries of the diagrams. For instance, if the color factors change
  by sign under the symmetry, then the numerator should also change by
  the same sign.
\item \textit {Unitarity cuts.}  The cuts of the ansatz should
  correctly match a spanning set of unitarity cuts.  This is simply a
  requirement that the ansatz be a correct representation of the
  Yang-Mills amplitude.
\item \textit {BCJ double-copy property}.  The double-copy property
  (\ref{ColorSubstitution}) should hold, allowing us to obtain
  amplitudes in the double-copy theory from gauge-theory ones simply
  by replacing color factors with kinematic numerators.
\end{enumerate}
Except for properties 3 and 7, these are standard properties that ordinary
Feynman diagrams in Feynman gauge possess.  The ghost contributions of
Feynman diagrams ensure that physical-state projectors are properly
reconstructed across unitarity cuts, but they do so in a way that maintains
the manifest locality of the numerators.  We are of course seeking a
representation of the amplitude with all of these properties as well
as additional ones, in particular the nontrivial final property 7 that
the double-copy property holds.  Note that we have not imposed that
the global BCJ identities hold manifestly, only that the weaker
double-copy property holds.

For the two-loop four-point amplitude, the set of diagrams with only
cubic vertices is given by the independent relabelings of the diagrams in
\fig{Fig:TwoLoopDiags}.  As usual, tadpoles have a vanishing color
factor, and bubble-on-external-leg contributions are dropped because
they integrate to zero in dimensional regularization through a
cancellation of ultraviolet and infrared singularities.  (However, one
needs to be aware of this cancellation when trying to extract only an
ultraviolet divergence or an infrared divergence from the amplitude.)

We start by imposing the first five conditions on the ansatz.  
Each diagram numerator ansatz, $n_{j}$, is a linear combination of monomials, $M_{jk}$, subject
to the power-counting constraint in property 2 above:
\begin{equation}
n_j = \sum_k a_{jk} M_{jk}\,,
\end{equation}
where $a_{jk}$ are the undetermined coefficients of the ansatz for diagram $(j)$.
The monomials are built out of independent dot products,
\begin{equation}
\pol_i \cdot \pol_j\,, \hskip .6 cm 
\pol_i \cdot k_j\,, \hskip .6 cm 
\pol_i \cdot p\,, \hskip .6 cm 
\pol_i \cdot q\,, \hskip .6 cm 
k_i \cdot p\,, \hskip .6 cm
k_i \cdot q\,,  \hskip .6 cm
p^2\,,  \hskip .6 cm
q^2\,,  \hskip .6 cm
p \cdot q\,,  \hskip .6 cm
s\,, \hskip .6 cm
t\,,
\end{equation}
with leg labels $i,j = 1,2,3,4$.  We only include monomials that are
independent under momentum-conservation, on-shell and
transversality conditions:
\begin{equation}
k_{4} = - k_{1} - k_{2} - k_{3}\,, \hskip .7 cm 
k_{i}^2 = 0\,, \hskip .7 cm
k_{i}\cdot \pol_{i} = 0\,.
\end{equation}
By imposing constraints 1--4, we obtain an ansatz with 9814 terms for
diagram (1), 9452 terms for diagram (2), 9902 terms for diagram (3)
and so on.  Due to property 5, many terms are related by diagram
symmetries.  Imposing these symmetry relations reduces the number of
undetermined coefficients respectively to 2703, 4748, 2546 and so on for these
diagrams.

\subsection{Global BCJ Identities on Integrand}
\label{sec:GlobalBCJ}

Can we consistently impose the global BCJ identities with the above
constraints?  If we could, then this would be an efficient route to
constructing the loop integrand. In attempting this, we follow the
strategy described in detail in Ref.~\cite{ck4l}. By imposing global
BCJ identities, we are strengthening constraint 7 to 
\begin{enumerate} 
\item[$7'\!.$] \textit{BCJ duality manifest in integrand.} We
demand that the diagram numerators obey the full set of 
global dual Jacobi relations in \eqn{BCJDuality}.
\end{enumerate}
This constraint allows us to express all numerators in terms of two
master diagrams.  For the two-loop four-point amplitudes, we choose
the masters to be diagrams (1) and (2) in \fig{Fig:TwoLoopDiags}.
These are then the only diagrams that require an ansatz, and the
values of the remaining diagrams are generated by the BCJ numerator
relations.  This gives us an ansatz for the entire amplitude.  As
mentioned, the two master diagrams start with 9814 and 9452 terms,
respectively, where each term is multiplied by a free parameter.
%
%
With the first five constraints and global BCJ constraints imposed, we
are reduced to 1279 total free parameters, noting that
diagram-symmetry constraints fix a large number of coefficients. We
then systematically step through and impose the spanning set of
unitarity cuts in \fig{Fig:UnitarityCuts}.

The standard way to evaluate the cuts is to sew together tree
amplitudes appearing in the cuts using physical-state projectors.  The
$D$-dimensional projectors introduce light-cone denominators, which
must cancel away in the final gauge-invariant cut expressions.  We
avoid introducing such spurious denominators by using ghost
fields via Feynman diagrams.  While a Feynman-diagram representation
is not particularly enlightening or compact, it gives a simple means
to generate target expressions for checking unitarity cuts.  The cuts
of ordinary Feynman-gauge Feynman diagrams automatically contain only
the usual Feynman propagators and therefore never introduce spurious
denominators in the first place.  In any case, the unitarity cuts are
gauge invariant, and their values do not depend on how they are
generated.

Once target expressions for the cuts are generated, the next step is
to fix some or all of the remaining parameters in the ansatz by
matching its cuts to the target cuts.  After some effort, we find that
the result of this procedure is that the constraints 1--6 listed above
are incompatible with finding a solution to the global BCJ
identities. We can consistently impose unitarity cuts (b) and (c) of
\fig{Fig:UnitarityCuts}, but the iterated two-particle cut (a) by
itself is inconsistent with conditions 1--6 and $7'$. We can localize
the problem by observing that unitarity cuts (b) and (c) of
\fig{Fig:UnitarityCuts} are simultaneously consistent and fix all
relevant contact terms in the ansatz except those proportional to
$(p+q)^{2}$, corresponding to the labels in
\fig{Fig:TwoLoopDiags}. Hence, the contact terms involving the middle
propagator of diagrams (1), (4) and (8) in \fig{Fig:TwoLoopDiags}
cannot be consistently constrained. (We directly confirmed that this
is the source of the problem by using a cut-merging
procedure~\cite{CutMerge} through which we added terms to these
diagrams proportional to $(p+q)^2$, making them consistent with
unitarity at the cost of losing the double-copy property.)

One could hope to conquer this obstacle by relaxing some or all of
constraints 1--4. In simple cases, this is usually the best strategy.
However, the minimal ansatz for each diagram numerator is already
fairly complicated, and the complexity grows rapidly as these
constraints are released.  This becomes more problematic as the number
of loops increases.  The difficulty here in finding an ansatz
compatible with both unitarity and global BCJ constraints is a simpler
version of difficulties encountered at five loops in $\NeqFour$
super-Yang-Mills theory~\cite{FiveLoopN4YM}.

\subsection{BCJ Duality on Generalized Unitarity Cuts}

Here we explore an alternative tactic of preserving constraints 
1--5 that keeps the complexity of each numerator ansatz under control,
but instead releases the constraint that global BCJ duality is manifest
in the integrand.  We do so in such a way that preserves the crucial
double-copy property that allows us to obtain gravity integrands from
gauge-theory ones.  This strategy amounts to a trade-off between
maintaining the relative simplicity of minimal ans\"atze and losing
the global BCJ relations that allow us to write the full amplitude in
terms of master diagrams.

In order to maintain the double-copy property 7, we impose instead
the following condition:
\begin{enumerate}
\item [$7''\!\!.$] \textit{BCJ duality on generalized unitarity cuts.}  Demand that
  BCJ duality is manifest in a spanning set of generalized
  unitarity cuts of the amplitude.
\end{enumerate}
This effectively means that the cuts of amplitudes are expressed in
terms of sums of products of tree amplitudes, where each tree is in a
form where BCJ duality is manifest. Tree amplitudes can always be put
into a BCJ-satisfying form, but what makes this condition nontrivial is the
requirement of a single integrand having the property that for each cut in a
spanning set of cuts, BCJ duality is manifest without needing
rearrangements.  This then guarantees that under the replacement
(\ref{ColorSubstitution}) in the integrand, the cuts match those
of the double-copy gravity theory.  It is however a weaker condition
than requiring that BCJ duality be manifest globally in the integrand itself.

To find an integrand with the desired properties, we use a
distinct ansatz for each diagram in \fig{Fig:TwoLoopDiags}.
Applying conditions 1--4 above, we start with a rather large
 ansatz containing 120904 parameters.  This is
reduced to 28204 free parameters by condition 5, which requires
that the numerators respect the diagram symmetries.

As in \sect{sec:GlobalBCJ}, we compare each unitarity cut against the
target cuts.  In addition, we impose that the numerators of the cut
diagrams obey all BCJ relations, as illustrated in the examples of
\fig{Fig:TwoLoopCutBCJ}. In this case, there is no problem finding a
solution to both the BCJ relations and the spanning set of cuts in
\fig{Fig:UnitarityCuts}.  In fact, the final solution has 6322 free
parameters.  These free parameters do not alter any unitarity cuts,
but merely move contact terms between diagrams while maintaining 
BCJ duality on the spanning set of cuts in \fig{Fig:UnitarityCuts}.

\begin{figure}[tb]
\includegraphics[scale=.45]{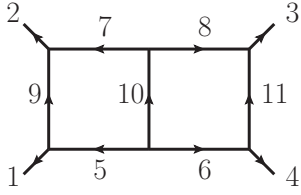}
\caption{A diagram with labels corresponding to those of the ancillary file.
The arrows give the direction of the momenta.}
\label{Fig:FullLabels}
\end{figure}

The solution is lengthy and is found in the ancillary
file~\cite{AncFile}\footnote{The remaining 6322 free coefficients are
  set to zero in this file.  It would be interesting to see if there
  are other useful properties that could be enforced with wise choices
  of the free parameters.}.  Its length is largely due to the fact
that we use formal polarization vectors, which allows us the
generality to work in $D$ dimensions.  In the ancillary file,
\code{numerators.m}, the vertices, inverse propagators, numerators,
color factors and symmetry factors of each diagram in
\Fig{Fig:TwoLoopDiags} are given in \code{Mathematica} syntax as a
list. In
this syntax, the information of diagram $(j)$ needed for
\eqns{CubicRepresentationLoop}{DoubleCopyLoop} can be accessed by
\code{vertices[[j]]}, \code{propagators[[j]]}, \code{numerators[[j]]},
\code{color[[j]]} and \code{symmetry[[j]]}, where ``\code{[[j]]}''
takes the $j$th entry in the list.  For example, the planar
double box, diagram (1), is accessed by \code{vertices[[1]]}, which
returns
\begin{align}
& \code{\{\{-1\},\{-2\},\{-3\},\{-4\},\{1,9,-5\},\{2,-7,-9\},}
\nonumber \\
&\hspace{1cm}\code{\{3,-11,-8\},\{4,-6,11\},\{5,10,6\},\{7,8,-10\}\}}\,\,.
\end{align}
The structure of \code{vertices[[j]]} is such that a one-element list
represents an external momentum label, a three-element list represents
a cubic vertex in clockwise ordering, and a positive (negative) number
represents an outgoing (incoming) momentum from (to) the vertex. These
labels match those displayed in \fig{Fig:FullLabels} for the planar
double box.

Considering diagram (1) as an example, the inverse propagators,
denoted $\prod_{\alpha_{1}}p^{2}_{\alpha_{1}}$ in
\eqns{CubicRepresentationLoop}{DoubleCopyLoop}, are accessed by,
\code{propagators[[1]]}, which returns
\begin{align}
& 
\code{(k[1]-k[5])\caret 2*(k[1]+k[2]-k[5])\caret 2*k[5]\caret 2*k[6]\caret 2*}
\nonumber \\
&\hspace{1cm}\code{(k[1]+k[2]+k[6])\caret 2*(k[1]+k[2]+k[3]+k[6])\caret 2*}
\nonumber \\
&\hspace{1cm}\code{(k[5]+k[6])\caret 2}\,\,.
\end{align}
These of course follow from momentum conservation.  The corresponding
numerator for diagram (1), $n_{1}$, is accessed by
\code{numerators[[1]]}, which returns
\begin{align}
& \code{(I/189)*(-140278+24959*Ds)*ke[1,2]*ke[2,1]*}
\nonumber \\
&\hspace{1cm}\code{ke[2,3]*ke[2,4]*kk[1,2] + $\cdots$}\,\,,
\end{align}
where the `$\cdots$' signifies 8223 additional terms found in the
ancillary file.  Note that we group factors of $i$ coming from the
Feynman propagators with the numerator. Also, take note of the overall
prefactors in \eqns{CubicRepresentationLoop}{DoubleCopyLoop}.  The
parameter \code{Ds} is a state counting parameter that comes from the
contraction of metric tensors. It is the same parameter as $D_s$ in
\eqn{eq:PNP}. Depending on the regularization scheme, in four
dimensions it can be $\code{Ds} = 4$ or $\code{Ds} = 4-2\eps$.

The color factor can be read off directly from the vertices by
replacing each triplet of numbers in the vertices $\{i,j,k\}$ with an
$\tilde{f}^{a_i a_j a_k}$  (dropping the index signs).  For example,
\code{color[[1]]} returns
\begin{align}
& \code{tf[1,9,5]*tf[2,7,9]*tf[3,11,8]*} \nonumber \\
&\hspace{1cm} \code{tf[4,6,11]*tf[5,10,6]*tf[7,8,10]} \,,
\end{align}
corresponding to the diagram (1) color factor,
\begin{align}
c_1 = \tilde{f}^{a_{1}a_{9}a_{5}}
\tilde{f}^{a_{2}a_{7}a_{9}}
\tilde{f}^{a_{3}a_{11}a_{8}}
\tilde{f}^{a_{4}a_{6}a_{11}}
\tilde{f}^{a_{5}a_{10}a_{6}}
\tilde{f}^{a_{7}a_{8}a_{10}}\,.
\end{align}
Finally, the symmetry factor $S_{1}$ for diagram (1), for instance, is
given by \code{symmetry[[1]]}, which returns 4.

Taking the numerators, propagators, color and symmetry factors in the
ancillary file and inserting them into \eqn{CubicRepresentationLoop}
gives the two-loop four-point Yang-Mills amplitude.  Similarly, the
corresponding double-copy gravity amplitude is obtained by inserting
the appropriate factors into \eqn{DoubleCopyLoop}. We have explicitly
checked that the constructed gauge and gravity amplitudes correctly
satisfy the spanning set of generalized unitarity cuts given in
\fig{Fig:UnitarityCuts}. In the gravity amplitude verification, the
integrands were constructed in two different ways as a cross check:
The primary way uses two copies of the gauge-theory numerators from the
ancillary file, whereas the second way uses one copy from the ancillary
file and the other copy from numerators generated by nonsupersymmetric
Yang-Mills theory Feynman-gauge Feynman rules. Both ways give
correct gravity results compatible with the unitarity constraints.

\section{Ultraviolet Properties}
\label{sec:Ultraviolet}

In this short section, we summarize results on the ultraviolet
singularities derived from the identical-helicity four-graviton
amplitude discussed in \sect{sec:AllPlus}.  The counterterm is the
well-studied two-loop $R^3$
counterterm~\cite{GoroffSagnotti,Evanescent}.  These results summarize
those of Ref.~\cite{nonSUSYBCJ} for the bare contribution and
Refs.~\cite{Evanescent} and \cite{LongPaper} for the subdivergence
subtractions.

The result for the two-loop bare divergence is already given 
in  Ref.~\cite{nonSUSYBCJ}.
That paper uses a different
representation of the integrand, where global BCJ relations hold
between the diagrams. In any case, after integrating we obtain the
same bare ultraviolet divergence.  Integral tables may be found in the
appendices of Ref.~\cite{nonSUSYBCJ}. Using dimensional regularization,
the infrared and ultraviolet divergences are mixed together.  However,
the infrared singularities are simple and known ahead of
time~\cite{IRPapers}, so they are easily subtracted out.
(Ref.~\cite{nonSUSYBCJ} also used an alternate method based on
introducing a mass regulator; the results for the ultraviolet
divergences are the same for either method.)  The net result is that
the bare two-loop divergence is
\begin{equation}
\left.\mathcal{M}^{(2)}(1^+,2^+,3^+,4^+)\right|_{\mathrm{bare~UV~div.}} 
=\frac{1}{\epsilon} \, \frac{83}{2700} \K\,,
\label{eq:twoLoopUV}
\end{equation}
where
\begin{equation}
\K \equiv \Bigl(\frac{\kappa}{2} \Bigr)^6 \frac{i}{(4 \pi)^4}
s t u \biggl(\frac{[1 2] [3 4]}
{\langle 1 2 \rangle \langle 3 4 \rangle} \biggr)^2 \,.
\label{KDef}
\end{equation}

However, as discussed in Ref.~\cite{Evanescent}, an important subtlety
is that there are also subdivergence subtractions even though there
are no corresponding one-loop divergences in $D = 4$.  The double-copy
theory does have one-loop divergences~\cite{HooftVeltmanMatter,
  nonSUSYBCJ}, but these divergences are in four-matter amplitudes
which cannot appear as subdivergences in the two-loop four-graviton
amplitude.  The origin of this curious behavior is evanescent
operators, which can act as counterterms.  The Gauss-Bonnet term
is one such evanescent operator. In the double-copy theory,
the Gauss-Bonnet counterterm is
\begin{equation}
\mathcal{L}_{\mathrm{GB}}^{\rm CT} = -\frac{1}{(4 \pi)^2} \Bigl(\frac{53}{90} + 
   \frac{1}{360} + \frac{91}{360}  \Bigr) 
\sqrt{-g} (R^2 - 4 R_{\mu\nu}^2 + R_{\mu\nu\rho\sigma}^2)\,,
\label{GBCT}
\end{equation}
where the three numbers in the parenthesis correspond to the contributions
from the graviton, dilaton and antisymmetric tensor, respectively.
The coefficients are proportional to the ones that appear in the trace
anomaly~\cite{ConformalAnomaly}.  The Gauss-Bonnet theorem implies
that in four dimensions, the operator (\ref{GBCT}) is a total
derivative.  When antisymmetric tensors are coupled to gravity,
another relevant one-loop four-point divergence is that of two
gravitons and two antisymmetric tensors, which has a counterterm,
\begin{equation}
\LRHH^{\rm CT} = - \frac{5}{3}\left(\frac{\kappa}{2}\right)^2 
\frac{1}{(4\pi)^2} \frac{1}{\eps}
\sqrt{-g} R^{\mu\nu}_{\ \ \rho\sigma} H_{\mu\nu\alpha}
H^{\alpha\rho\sigma} \,.
\label{RHHCT}
\end{equation}
Like the Gauss-Bonnet counterterm
(\ref{GBCT}), this operator is evanescent.  In $D=4$, we can dualize
the antisymmetric tensors to scalars, which collapses the Riemann
tensor into the Ricci scalar and tensor.  Under field redefinitions,
they can then be removed in favor of the dualized scalars, eliminating
the one-loop divergence in two-graviton two-antisymmetric-tensor
amplitudes when the external states are in $D=4$.  

\begin{figure}[tb]
\begin{center}
\subfloat[]{\includegraphics[scale=.35]{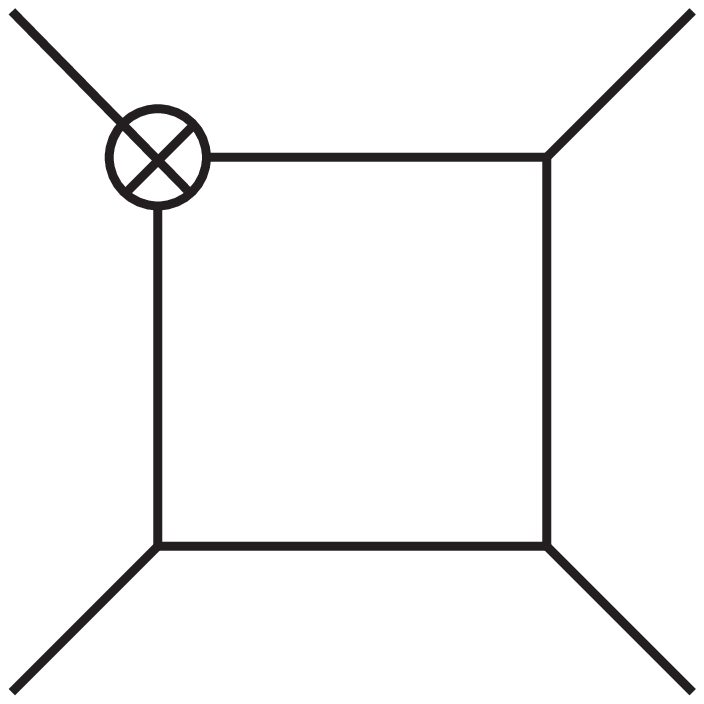}}
\hspace{1cm}
\subfloat[]{\raisebox{.55cm}{\includegraphics[scale=.35]{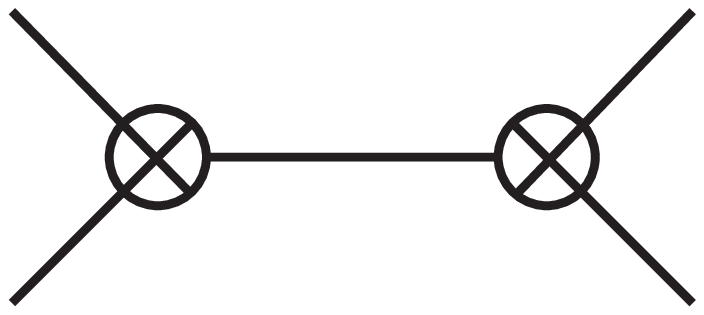}}}
\end{center}
\vskip -.2cm
\caption[a]{\small Representative diagrams of the 
(a) single-counterterm and (b) double-counterterm insertions.
\label{CounterSingleDoubleFigure}
}
\end{figure}

As discussed in Ref.~\cite{Evanescent}, there are two types of
counterterm subtractions, as illustrated by representative diagrams
in \fig{CounterSingleDoubleFigure}.  The single-counterterm insertion
subdivergence corresponding to \fig{CounterSingleDoubleFigure}(a) is~\cite{LongPaper}
\begin{equation}
\mathcal{M}^{(2)}(1^+,2^+,3^+,4^+)\Bigr|_{\mathrm{single~CT}} 
= \frac{1}{\epsilon} \, \Bigl(-\frac{6004}{675} + \frac{25}{3} \Bigr) \K \,,
\label{eq:SingleCTUV}
\end{equation}
where the first number is the contribution from the Gauss-Bonnet counterterm 
in \eqn{GBCT} and the second is the contribution from the $RH\!H$ counterterm in \eqn{RHHCT}.
In addition, there is the double Gauss-Bonnet counterterm insertion illustrated by 
a representative diagram in \fig{CounterSingleDoubleFigure}(b). This
is given by
\begin{equation}
\mathcal{M}^{(2)}(1^+,2^+,3^+,4^+)\Bigr|_{\mathrm{double~CT}}
= \frac{1}{\epsilon} \, \frac{11552}{675} \K \,.
\label{eq:DoubleCTUV}
\end{equation}

Adding together the three contributions in Eqs.~(\ref{eq:twoLoopUV}),
(\ref{eq:SingleCTUV}) and (\ref{eq:DoubleCTUV}) gives the total
two-loop divergence,
\begin{equation}
\mathcal{M}^{(2)}(1^+,2^+,3^+,4^+)\Bigr|_{\mathrm{total}}
= \frac{1}{\epsilon} \, \frac{199}{12}\K \,.
\label{eq:total}
\end{equation}
This divergence can be removed from the amplitude by adding
an $R^3$ counterterm to the theory,
\begin{equation}
\LRRR^{\mathrm{CT}} =
\frac{199}{720} \left(\frac{\kappa}{2}\right)^2 \frac{1}{(4 \pi)^4} \frac{1}{\eps}
 \sqrt{-g}\, R^{\alpha \beta}{\!}_{\gamma\delta}
R^{\gamma \delta}{\!}_{\rho\sigma} R^{\rho\sigma}{\!}_{\alpha\beta} \,.
\label{GSDiv}
\end{equation}
Up to the coefficient, this is the same divergence that appears in 
pure gravity at two loops~\cite{GoroffSagnotti, Evanescent}.  
As already noted, the double-copy theory also has a one-loop divergence
in the matter sector~\cite{HooftVeltmanMatter}.
                              
As discussed in Ref.~\cite{Evanescent}, the divergence itself is not
physical and is modified by duality transformations.  However, the
renormalization scale is unaltered.  The coefficient of the $\ln\mu^2$
dependence is easily extracted by noting that, for the bare two-loop
part, the $\ln\mu^2$ coefficient is twice the coefficient of the
$1/\eps$ divergence, for the single counterterm, it is equal to the
$1/\eps$ coefficient, and for the double-insertion tree contribution,
it vanishes.  This follows from dimensional analysis of the loop
integrals, with measure $\int d^{4-2\eps}\ell$ per loop, which requires an
overall factor of $\mu^{2L\eps}$ at $L$ loops.  The counterterm
subtractions themselves are pure poles which do not carry such factors.
This then gives the $\ln\mu^2$ dependence,
\begin{equation}
\mathcal{M}^{\twoloop}_4(1^+,2^+,3^+,4^+) \Bigr|_{\ln\mu^2} =  
- \K \,\frac{1}{2}\, \ln\mu^2\,,
\end{equation}
in agreement with the general formula given in Ref.~\cite{Evanescent},
\begin{equation}
\mathcal{M}^{\twoloop}_4(1^+,2^+,3^+,4^+) \Bigr|_{\ln\mu^2} = 
 - \K \,\frac{N_b-N_{\!f}}{8}\, \ln\mu^2\,,
\label{NbNf}
\end{equation}
where $N_b$ is the number of bosonic states and $N_{\! f}$ is the number of
fermionic states.

Ref.~\cite{LongPaper} will provide further details on a variety of
theories, including the double-copy gravity theory
(\ref{eq:LagrangianGDA}), as well as present full amplitudes including
their finite parts.  It is an interesting open problem to extract the
divergences from the amplitude in terms of formal polarization tensors
to ensure the consistency for all helicity configurations.  However,
one encounters high-rank tensor double-box integrals making the integration
nontrivial.

\section{Conclusions}
\label{sec:Conclusions}

Constructions of higher-loop amplitudes in supergravity have provided
a wealth of new nontrivial information on gravity theories.  This
includes tantalizing new effects such as enhanced ultraviolet
cancellations~\cite{N5FourLoop} and dependence of leading divergences
on evanescent effects and duality
transformations~\cite{Evanescent,LongPaper}.  In addition, the only two
known divergences in four dimensions in pure Einstein gravity and
pure ungauged supergravity theories
display anomaly-like behavior, as discussed in
Refs.~\cite{Evanescent,N4FourLoops}.  

To further explore these effects and to uncover new ones, we need more
efficient ways to obtain gravity amplitudes at high loop orders. At
present the most powerful means for doing so is based on the
duality between color and kinematics~\cite{BCJ,BCJLoop} in
conjunction with the unitarity method~\cite{UnitarityMethod}.
However, at sufficiently high loop orders or in complicated cases with
little or no supersymmetry, it can be nontrivial to find
representations of the amplitude where the duality is manifest.  It
may be that ans\"atze that are sufficiently general to be
compatible with both unitarity cuts and global BCJ constraints are
impractical to work with.  The strategy presented here for dealing
with such difficulties is to loosen the BCJ constraints by demanding
that they hold manifestly only on unitarity cuts instead of on uncut
integrands.  This allows us to use simpler ans\"atze for individual
diagrams yet retain the key double-copy property. The cost is that we
lose the ability to determine the gauge-theory integrand from a small
number of master diagrams.

After warming up on the case of identical helicities, we demonstrated
our strategy in action on the two-loop four-point pure Yang-Mills
amplitude with $D$-dimensional external polarizations.  We first showed that
a minimal ansatz, where locality, crossing symmetry and
Feynman-diagram-like power-counting are manifest
is not compatible simultaneously with both global BCJ duality
on the integrand and unitarity.  On the other hand,
no difficulties are encountered when the duality requirements are
loosened so that they are manifest only on a spanning set of
generalized unitarity cuts. The so-constructed Yang-Mills integrand
immediately produces a corresponding gravity integrand via the 
double-copy procedure.

Using results from Refs.~\cite{nonSUSYBCJ,Evanescent,LongPaper}, we
extracted the two-loop ultraviolet divergence from the
identical-helicity amplitude of the double-copy gravity theory.  As
explained in Ref.~\cite{Evanescent}, in contrast to the divergence
itself, the renormalization-scale dependence follows a universal
formula that depends only on the number of four-dimensional bosonic
and fermionic states and not on their spin. Further details will be
given in Ref.~\cite{LongPaper}.  It would be interesting to carry out
the same analysis on the case of general polarizations using
conventional dimensional regularization to confirm the generality of
the results.

We hope that the strategy presented here will help lead to new
constructions of multiloop (super)gravity amplitudes, and we look
forward to the new insights that they will provide.

\subsection*{Acknowledgments}

We thank John Joseph Carrasco, Clifford Cheung, Huan-Hang Chi, Tristan
Dennen, Lance Dixon, Henrik Johansson and Radu Roiban for many
interesting and useful discussions and collaboration on related work.
This material is based upon work supported by the Department of Energy
under Award Number DE-{S}C0009937. ZB is grateful to the Simons
Foundation for support.  We gratefully acknowledge Mani Bhaumik for
his continued generous support.


\end{document}